\documentclass[12pt,a4paper]{article}

\usepackage{ifthen} % for conditional statements
\newboolean{pdflatex}
\setboolean{pdflatex}{true} % False for eps figures 

\usepackage{booktabs}
\usepackage{multirow}
\usepackage{adjustbox}
\usepackage[abs]{overpic}
\usepackage[separate-uncertainty=true]{siunitx}
\usepackage{xcolor}
\usepackage{amsmath}
\usepackage{subcaption}

\newboolean{articletitles}
\setboolean{articletitles}{true} % False removes titles in references

\DeclareSIUnit\MeV{\mega\eV}
\DeclareSIUnit\GeV{\giga\eV}
\DeclareSIUnit\TeV{\tera\eV}
\DeclareSIUnit\PeV{\peta\eV}
\DeclareSIUnit\pernb{\per\nano\barn}
\DeclareSIUnit\perpb{\per\pico\barn}
\DeclareSIUnit\perfb{\per\femto\barn}

\DeclareSIUnit\clight{\text{\ensuremath{c}}}
\DeclareSIUnit[per-mode=symbol]\mevc{\MeV\per\clight}
\DeclareSIUnit[per-mode=symbol]\gevc{\GeV\per\clight}
\DeclareSIUnit[per-mode=symbol]\tevc{\TeV\per\clight}
\DeclareSIUnit[per-mode=symbol]\mevcc{\MeV\per\square\clight}
\DeclareSIUnit[per-mode=symbol]\gevcc{\GeV\per\square\clight}
\DeclareSIUnit[per-mode=symbol]\tevcc{\TeV\per\square\clight}
\newboolean{uprightparticles}
\setboolean{uprightparticles}{false} %True for upright particle symbols

\newboolean{inbibliography}
\setboolean{inbibliography}{false} %True once you enter the bibliography

\def\paperauthors{L.~Anderlini, F.~Archilli, A.~Cardini, V.~Cogoni, M.~Fontana, G.~Graziani, N.~Kazeev, H.~Kuindersma, R.~Oldeman, M.~Palutan, M.~Santimaria, B.~Sciascia, P.~De~Simone, R.~Vazquez~Gomez}
\def\paperasciititle{Muon identification for LHCb Run 3}
\def\papertitle{Muon identification for LHCb Run 3}
\def\paperkeywords{{High Energy Physics}, {LHCb}}
\def\papercopyright{\the\year\ CERN for the benefit of the LHCb collaboration}
\def\paperlicence{CC-BY-4.0 licence}
\def\paperlicenceurl{https://creativecommons.org/licenses/by/4.0/}

\newif\ifreview
\reviewfalse

\usepackage[top=1in, bottom=1.25in, left=1in, right=1in]{geometry}

\usepackage{microtype}
\usepackage{lineno}  % for line numbering during review
\usepackage{xspace} % To avoid problems with missing or double spaces after predefined symbols
\usepackage{caption} %these three command get the figure and table captions automatically small

\usepackage{graphicx}  % to include figures (can also use other packages)
\usepackage{color}
\usepackage{colortbl}
\graphicspath{{./figs/}} % Make Latex search fig subdir for figures
\usepackage[section]{placeins} % keeps figures and tables within a section

\usepackage{amsmath} % Adds a large collection of math symbols
\usepackage{amssymb}
\usepackage{amsfonts}
\usepackage{upgreek} % Adds in support for greek letters in roman typeset

\newcommand*\patchAmsMathEnvironmentForLineno[1]{%
\expandafter\let\csname old#1\expandafter\endcsname\csname #1\endcsname
\expandafter\let\csname oldend#1\expandafter\endcsname\csname
end#1\endcsname
 \renewenvironment{#1}%
   {\linenomath\csname old#1\endcsname}%
   {\csname oldend#1\endcsname\endlinenomath}%
}
\newcommand*\patchBothAmsMathEnvironmentsForLineno[1]{%
  \patchAmsMathEnvironmentForLineno{#1}%
  \patchAmsMathEnvironmentForLineno{#1*}%
}
\AtBeginDocument{%
\patchBothAmsMathEnvironmentsForLineno{equation}%
\patchBothAmsMathEnvironmentsForLineno{align}%
\patchBothAmsMathEnvironmentsForLineno{flalign}%
\patchBothAmsMathEnvironmentsForLineno{alignat}%
\patchBothAmsMathEnvironmentsForLineno{gather}%
\patchBothAmsMathEnvironmentsForLineno{multline}%
\patchBothAmsMathEnvironmentsForLineno{eqnarray}%
}

\usepackage{hyperxmp}

\usepackage[pdftex,
            pdfauthor={\paperauthors},
            pdftitle={\paperasciititle},
            pdfkeywords={\paperkeywords},
            pdfcopyright={Copyright (C) \papercopyright},
            pdflicenseurl={\paperlicenceurl}]{hyperref}

\usepackage[all]{hypcap}

\usepackage{xspace} 
\usepackage{upgreek}

\def\lhcb {\mbox{LHCb}\xspace}

\def\lhc    {\mbox{LHC}\xspace}

\def\ecal   {ECAL\xspace}
\def\hcal   {HCAL\xspace}
\def\MagUp {\mbox{\em Mag\kern -0.05em Up}\xspace}

\def\lone   {L0\xspace}

\def\hltone {HLT1\xspace}
\def\hlttwo {HLT2\xspace}

\ifthenelse{\boolean{uprightparticles}}%
{
 
 \def\Pgamma      {\ensuremath{\upgamma}\xspace}

 \def\Pmu         {\ensuremath{\upmu}\xspace}                 
 \def\Pnu         {\ensuremath{\upnu}\xspace}                 
                  
 \def\Ppi         {\ensuremath{\uppi}\xspace}                 
                  
 \def\Prho        {\ensuremath{\uprho}\xspace}                 
                  
 \def\Ptau        {\ensuremath{\uptau}\xspace}

 \def\Ppsi        {\ensuremath{\uppsi}\xspace}

 \def\PDelta      {\ensuremath{\Delta}\xspace}                 
 \def\PXi      {\ensuremath{\Xi}\xspace}                 
 \def\PLambda      {\ensuremath{\Lambda}\xspace}                 
 \def\PSigma      {\ensuremath{\Sigma}\xspace}                 
 \def\POmega      {\ensuremath{\Omega}\xspace}                 
 \def\PUpsilon      {\ensuremath{\Upsilon}\xspace}                 
 
 \def\PB      {\ensuremath{\mathrm{B}}\xspace}                 
                  
 \def\PD      {\ensuremath{\mathrm{D}}\xspace}

 \def\PJ      {\ensuremath{\mathrm{J}}\xspace}                 
 \def\PK      {\ensuremath{\mathrm{K}}\xspace}

 \def\Pb      {\ensuremath{\mathrm{b}}\xspace}                 
 \def\Pc      {\ensuremath{\mathrm{c}}\xspace}

 \def\Pi      {\ensuremath{\mathrm{i}}\xspace}

 \def\Pp      {\ensuremath{\mathrm{p}}\xspace}

 \def\Ps      {\ensuremath{\mathrm{s}}\xspace}

}
{
 
 \def\Pgamma      {\ensuremath{\gamma}\xspace}

 \def\Pmu         {\ensuremath{\mu}\xspace}                 
 \def\Pnu         {\ensuremath{\nu}\xspace}                 
                  
 \def\Ppi         {\ensuremath{\pi}\xspace}                 
                  
 \def\Prho        {\ensuremath{\rho}\xspace}                 
                  
 \def\Ptau        {\ensuremath{\tau}\xspace}

 \def\Ppsi        {\ensuremath{\psi}\xspace}                 
                  
 \mathchardef\PDelta="7101
 \mathchardef\PXi="7104
 \mathchardef\PLambda="7103
 \mathchardef\PSigma="7106
 \mathchardef\POmega="710A
 \mathchardef\PUpsilon="7107
                  
 \def\PB      {\ensuremath{B}\xspace}                 
                  
 \def\PD      {\ensuremath{D}\xspace}

 \def\PJ      {\ensuremath{J}\xspace}                 
 \def\PK      {\ensuremath{K}\xspace}

 \def\Pb      {\ensuremath{b}\xspace}                 
 \def\Pc      {\ensuremath{c}\xspace}

 \def\Pi      {\ensuremath{i}\xspace}

 \def\Pp      {\ensuremath{p}\xspace}

 \def\Ps      {\ensuremath{s}\xspace}

}

\makeatletter
\ifcase \@ptsize \relax% 10pt
  \newcommand{\miniscule}{\@setfontsize\miniscule{4}{5}}% \tiny: 5/6
\or% 11pt
  \newcommand{\miniscule}{\@setfontsize\miniscule{5}{6}}% \tiny: 6/7
\or% 12pt
  \newcommand{\miniscule}{\@setfontsize\miniscule{5}{6}}% \tiny: 6/7
\fi
\makeatother

\DeclareRobustCommand{\optbar}[1]{\shortstack{{\miniscule (\rule[.5ex]{1.25em}{.18mm})}
  \\ [-.7ex] $#1$}}

\def\mun        {{\ensuremath{\Pmu^-}}\xspace} % muon negative (\mum is taken)

\def\taum       {{\ensuremath{\Ptau^-}}\xspace}

\def\neu        {{\ensuremath{\Pnu}}\xspace}

\def\neum       {{\ensuremath{\neu_\mu}}\xspace}

\def\neut       {{\ensuremath{\neu_\tau}}\xspace}

\def\g      {{\ensuremath{\Pgamma}}\xspace}

\def\squark    {{\ensuremath{\Ps}}\xspace}

\def\cquark    {{\ensuremath{\Pc}}\xspace}

\def\bquark    {{\ensuremath{\Pb}}\xspace}

\def\pion   {{\ensuremath{\Ppi}}\xspace}
\def\piz    {{\ensuremath{\pion^0}}\xspace}

\def\pip    {{\ensuremath{\pion^+}}\xspace}
\def\pim    {{\ensuremath{\pion^-}}\xspace}

\def\rhomeson {{\ensuremath{\Prho}}\xspace}

\def\rhop     {{\ensuremath{\rhomeson^+}}\xspace}

\def\kaon    {{\ensuremath{\PK}}\xspace}
  \def\Kbar    {{\kern 0.2em\overline{\kern -0.2em \PK}{}}\xspace}

\def\KorKbar    {\kern 0.18em\optbar{\kern -0.18em K}{}\xspace}

\def\Kp      {{\ensuremath{\kaon^+}}\xspace}
\def\Km      {{\ensuremath{\kaon^-}}\xspace}

  \def\Dbar    {{\kern 0.2em\overline{\kern -0.2em \PD}{}}\xspace}
\def\D       {{\ensuremath{\PD}}\xspace}

\def\DorDbar    {\kern 0.18em\optbar{\kern -0.18em D}{}\xspace}

\def\Ds      {{\ensuremath{\D^+_\squark}}\xspace}
\def\Dsp     {{\ensuremath{\D^+_\squark}}\xspace}

\def\Dss     {{\ensuremath{\D^{*+}_\squark}}\xspace}

\def\B       {{\ensuremath{\PB}}\xspace}
\def\Bbar    {{\ensuremath{\kern 0.18em\overline{\kern -0.18em \PB}{}}}\xspace}

\def\BorBbar    {\kern 0.18em\optbar{\kern -0.18em B}{}\xspace}

\def\Bu      {{\ensuremath{\B^+}}\xspace}

\def\Bp      {{\ensuremath{\Bu}}\xspace}

\def\Bd      {{\ensuremath{\B^0}}\xspace}
\def\Bs      {{\ensuremath{\B^0_\squark}}\xspace}

\def\jpsi     {{\ensuremath{{\PJ\mskip -3mu/\mskip -2mu\Ppsi\mskip 2mu}}}\xspace}

  \def\Y#1S{\ensuremath{\PUpsilon{(#1S)}}\xspace}% no space before {...}!

\def\proton      {{\ensuremath{\Pp}}\xspace}

\def\Lz          {{\ensuremath{\PLambda}}\xspace}
\def\Lbar        {{\ensuremath{\kern 0.1em\overline{\kern -0.1em\PLambda}}}\xspace}
\def\LorLbar    {\kern 0.18em\optbar{\kern -0.18em \PLambda}{}\xspace}

\def\Lb      {{\ensuremath{\Lz^0_\bquark}}\xspace}

\def\Lc      {{\ensuremath{\Lz^+_\cquark}}\xspace}

\newcommand{\decay}[2]{\ensuremath{#1\!\to #2}\xspace}         % {\Pa}{\Pb \Pc}

\def\to                 {\ensuremath{\rightarrow}\xspace}

\def\AT#1     {\ensuremath{A_{\mathrm{T}}^{#1}}\xspace}           % 2

\def\C#1      {\ensuremath{\mathcal{C}_{#1}}\xspace}                       % 9
\def\Cp#1     {\ensuremath{\mathcal{C}_{#1}^{'}}\xspace}                    % 7
\def\Ceff#1   {\ensuremath{\mathcal{C}_{#1}^{\mathrm{(eff)}}}\xspace}        % 9  
\def\Cpeff#1  {\ensuremath{\mathcal{C}_{#1}^{'\mathrm{(eff)}}}\xspace}       % 7
\def\Ope#1    {\ensuremath{\mathcal{O}_{#1}}\xspace}                       % 2
\def\Opep#1   {\ensuremath{\mathcal{O}_{#1}^{'}}\xspace}                    % 7

\newcommand{\tev}{\ifthenelse{\boolean{inbibliography}}{\ensuremath{~T\kern -0.05em eV}}{\ensuremath{\mathrm{\,Te\kern -0.1em V}}}\xspace}
\newcommand{\gev}{\ensuremath{\mathrm{\,Ge\kern -0.1em V}}\xspace}
\newcommand{\mev}{\ensuremath{\mathrm{\,Me\kern -0.1em V}}\xspace}
\newcommand{\kev}{\ensuremath{\mathrm{\,ke\kern -0.1em V}}\xspace}
\newcommand{\ev}{\ensuremath{\mathrm{\,e\kern -0.1em V}}\xspace}
\newcommand{\gevc}{\ensuremath{{\mathrm{\,Ge\kern -0.1em V\!/}c}}\xspace}
\newcommand{\mevc}{\ensuremath{{\mathrm{\,Me\kern -0.1em V\!/}c}}\xspace}
\newcommand{\gevcc}{\ensuremath{{\mathrm{\,Ge\kern -0.1em V\!/}c^2}}\xspace}
\newcommand{\gevgevcccc}{\ensuremath{{\mathrm{\,Ge\kern -0.1em V^2\!/}c^4}}\xspace}
\newcommand{\mevcc}{\ensuremath{{\mathrm{\,Me\kern -0.1em V\!/}c^2}}\xspace}

\def\mum  {\ensuremath{{\,\upmu\mathrm{m}}}\xspace}

\def\barn{\ensuremath{\mathrm{ \,b}}\xspace}

\def\invpb {\ensuremath{\mbox{\,pb}^{-1}}\xspace}

\def\mhz  {\ensuremath{{\mathrm{ \,MHz}}}\xspace}

\newcommand{\chisq}{\ensuremath{\chi^2}\xspace}

\def\gsim{{~\raise.15em\hbox{$>$}\kern-.85em
          \lower.35em\hbox{$\sim$}~}\xspace}
\def\lsim{{~\raise.15em\hbox{$<$}\kern-.85em
          \lower.35em\hbox{$\sim$}~}\xspace}

\def\sPlot{\mbox{\em sPlot}\xspace}

\def\pt         {\mbox{$p_{\mathrm{ T}}$}\xspace}

\def\tell1  {TELL1\xspace}
\def\ukl1   {UKL1\xspace}

\input{local-symbols-def}
\usepackage{cite} % Allows for ranges in citations
\usepackage{mciteplus}

\usepackage{longtable}

\begin{document}
%%%%%%%%%%%%%%%%%%%%%%%%%%%%%%%%%%%%%%%%%%%%%%%%%%

%%%%%%%%%%%%%%%%%%%%%%%%%
%%%%% Title     %%%%%%%%%
%%%%%%%%%%%%%%%%%%%%%%%%%
\renewcommand{\thefootnote}{\fnsymbol{footnote}}
\setcounter{footnote}{1}
\begin{titlepage}

\vspace*{-1.5cm}
\centerline{\large EUROPEAN ORGANIZATION FOR NUCLEAR RESEARCH (CERN)}
\vspace*{1.5cm}

\noindent
\begin{tabular*}{\linewidth}{lc@{\extracolsep{\fill}}r@{\extracolsep{0pt}}}
\ifthenelse{\boolean{pdflatex}}% Logo format choice
{\vspace*{-1.5cm}\mbox{\!\!\!\includegraphics[width=.14\textwidth]{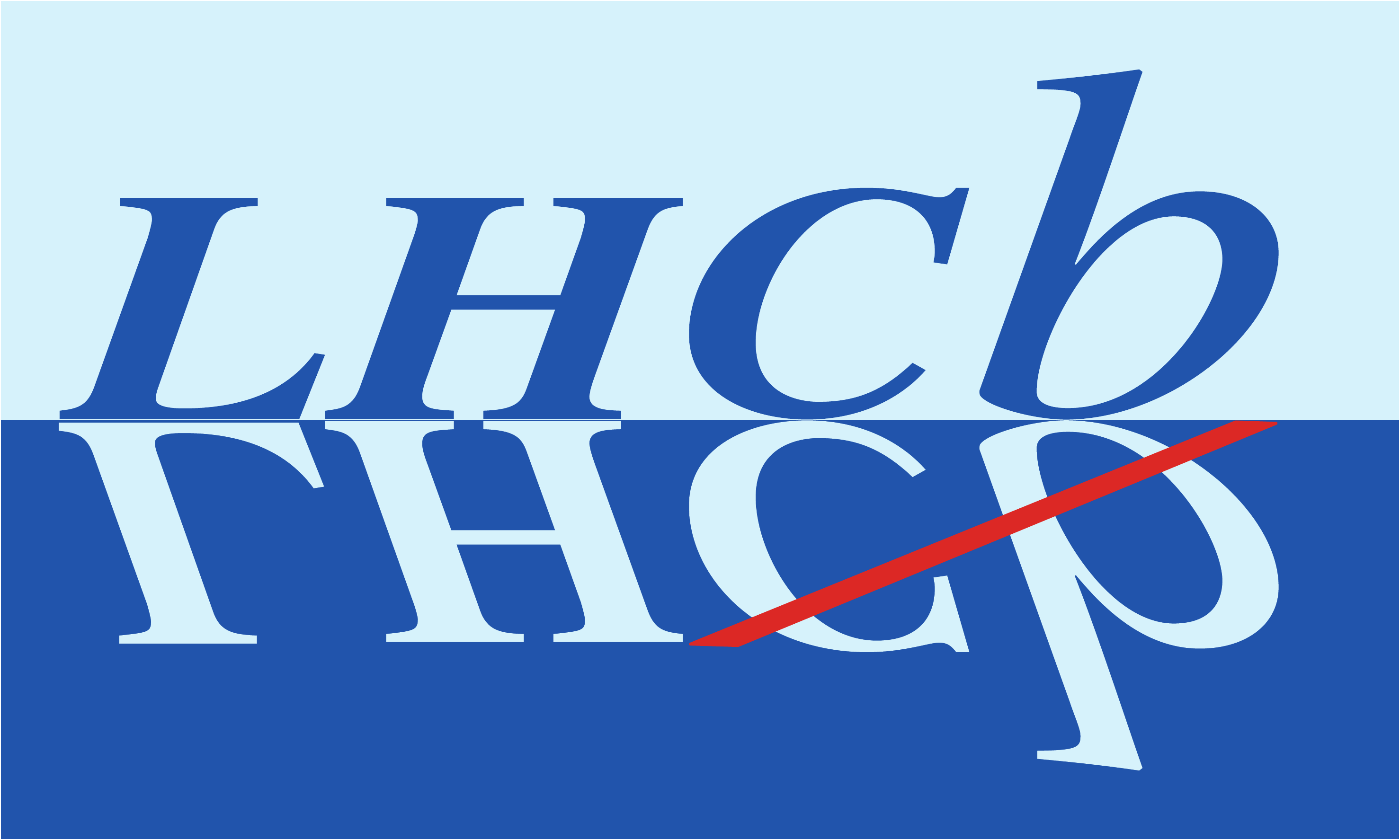}} & &}%
{\vspace*{-1.2cm}\mbox{\!\!\!\includegraphics[width=.12\textwidth]{figs/lhcb-logo.eps}} & &}%
\\
 & & CERN-LHCb-DP-2020-002 \\  % ID 
 & & \today \\ 
 & & \\
% not in paper \hline
\end{tabular*}

\vspace*{4.0cm}

% Title --------------------------------------------------
{\normalfont\bfseries\boldmath\huge
\begin{center}
  \papertitle 
\end{center}
}

\vspace*{2.0cm}

% Authors -------------------------------------------------
\begin{center}
L.~Anderlini$^1$, F.~Archilli$^2$, A.~Cardini$^3$, V.~Cogoni$^3$, M.~Fontana$^4$, G.~Graziani$^1$, N.~Kazeev$^{5,6}$, H.~Kuindersma$^7$, R.~Oldeman$^3$, M.~Palutan$^8$, M.~Santimaria$^8$, B.~Sciascia$^8$, P.~De~Simone$^8$, R.~Vazquez~Gomez$^{8,9}$
\bigskip\\
{\normalfont\itshape\footnotesize
$ ^1$Sezione di Firenze, INFN, Italy\\
$ ^2$Physikalisches Institut, Ruprecht-Karls-Universitat Heidelberg, Germany \\
$ ^3$Sezione di Cagliari, INFN, Italy\\
$ ^4$CERN, Switzerland\\
$ ^5$National Research University Higher School of Economics, Russia\\
$ ^6$The Sapienza University of Rome, Italy\\
$ ^7$Nikhef, National Institute for Subatomic Physics, Netherlands\\
$ ^8$Laboratori Nazionali di Frascati, INFN, Italy\\
$ ^9$IGFAE, Universidade de Santiago de Compostela, Spain\\
}
\end{center}
\vspace{\fill}

% Abstract -----------------------------------------------
\begin{abstract}
  \noindent
Muon identification is of paramount importance for the physics programme of LHCb. 
In the upgrade phase, starting from Run 3 of the LHC, the trigger of the experiment will be solely based on software.
The luminosity increase to $2\times10^{33}$\,cm$^{-2}$s$^{-1}$ will require an improvement of the muon identification criteria,
aiming at performances equal or better than those of Run 2, but in a much more challenging environment.
In this paper, two new muon identification algorithms developed in view of the \lhcb upgrade are presented, and 
their performance in terms of signal efficiency versus background reduction is shown. 
\end{abstract}
\vspace*{2.0cm}

\begin{center}
  Published in JINST 15 (2020) T12005
\end{center}

\vspace{\fill}

{\footnotesize 
% Edit macro in main.tex to keep metadata correct
\centerline{\copyright~\papercopyright. \href{\paperlicenceurl}{\paperlicence}.}}
\vspace*{2mm}

\end{titlepage}
\pagestyle{empty}

%%%%%%%%%%%%%%%%%%%%%%%%%%%%%%%%
%%%%%  EOD OF TITLE PAGE  %%%%%%
%%%%%%%%%%%%%%%%%%%%%%%%%%%%%%%%
\newpage
\setcounter{page}{2}
\mbox{~}

\renewcommand{\thefootnote}{\arabic{footnote}}
\setcounter{footnote}{0}

%%%%%%%%%%%%%%%%%%%%%%%%%%%%%%%%
%%%%%  Table of Content   %%%%%%
%%%%%%%%%%%%%%%%%%%%%%%%%%%%%%%%
%\tableofcontents
\cleardoublepage

%%%%%%%%%%%%%%%%%%%%%%%%%
%%%%% Main text %%%%%%%%%
%%%%%%%%%%%%%%%%%%%%%%%%%
\pagestyle{plain}
\setcounter{page}{1}
\pagenumbering{arabic}
\ifreview
  \linenumbers
  \clearpage
\else
\fi

\section{Introduction}
\label{sec:Introduction}

The \lhcb experiment~\cite{LHCb-DP-2014-002} at the \lhc is a single-arm forward spectrometer specialised in studying particles containing $b$ or $c$ quarks. Thanks to a versatile reconstruction and trigger system, the \lhcb physics programme has been extended to electroweak, soft QCD and even heavy-ion physics.
Many of the physics channels are identified by their very clean muon signatures, therefore muon identification and trigger are crucial for the success of the experiment.

\begin{figure}[b!]
    \centering
    \includegraphics[width=0.6\textwidth]{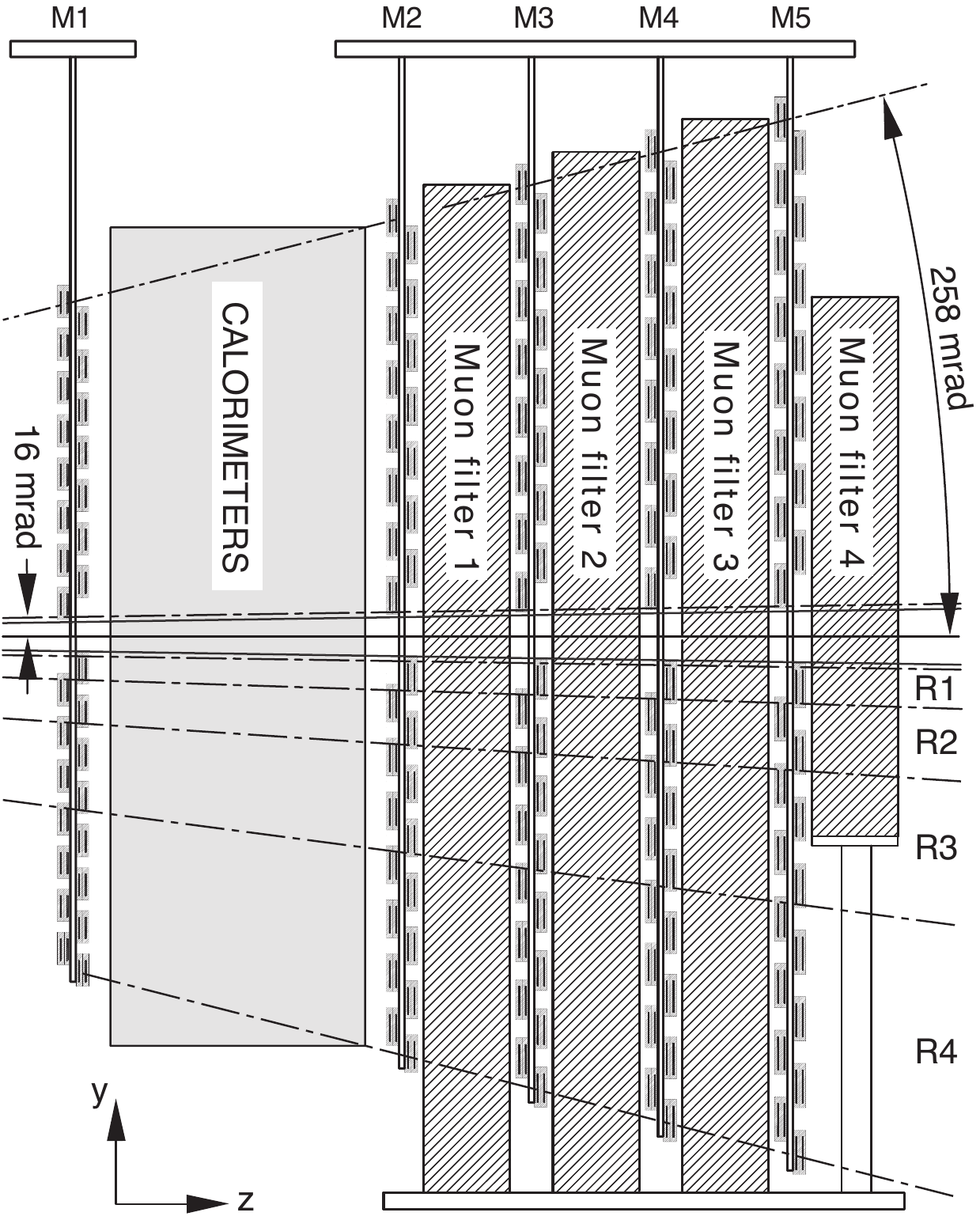}
    \caption{Side view of the muon system in the $y-z$ plane~\cite{LHCb-TDR-004}.}
    \label{fig:muon_system}
\end{figure}

A brief description of the Run 2 muon detector and reconstruction techniques follows, which sets the basis for the improvements later discussed in view of Run 3. A comprehensive description of the \lhcb trigger for Run 2 can be found in~\cite{Aaij_2019}. During Run 1 and Run 2, the tracking system of \lhcb provided a measurement of the momentum $(p)$ of charged particles with a relative uncertainty that varied from 0.5\% at low momentum to 1.0\% at 200\gevc~\cite{Aaij:2014pwa,dArgent:2017gpi}. The minimum distance of a track to a primary $pp$ collision vertex (PV), the impact parameter (IP), was measured with a resolution of $(15+29/\pt)\,\mum$, where
\pt is the component of the momentum transverse to the beam, in \gevc~\cite{LHCbVELOGroup:2014uea}. Muons were identified and triggered by a system composed of five stations, M1-M5, of rectangular shape, placed along the beam axis as shown in \Figref{fig:muon_system}.
Stations M2 to M5 were placed downstream the calorimeters, and were interleaved with $80$~cm-thick iron absorbers to select penetrating muons. The M1 station was placed in front of the calorimeters and used to improve the \pt measurement in the trigger~\cite{Aaij:2253050}. 

\begin{figure}[tb]
    \centering
    \includegraphics[width=0.6\textwidth]{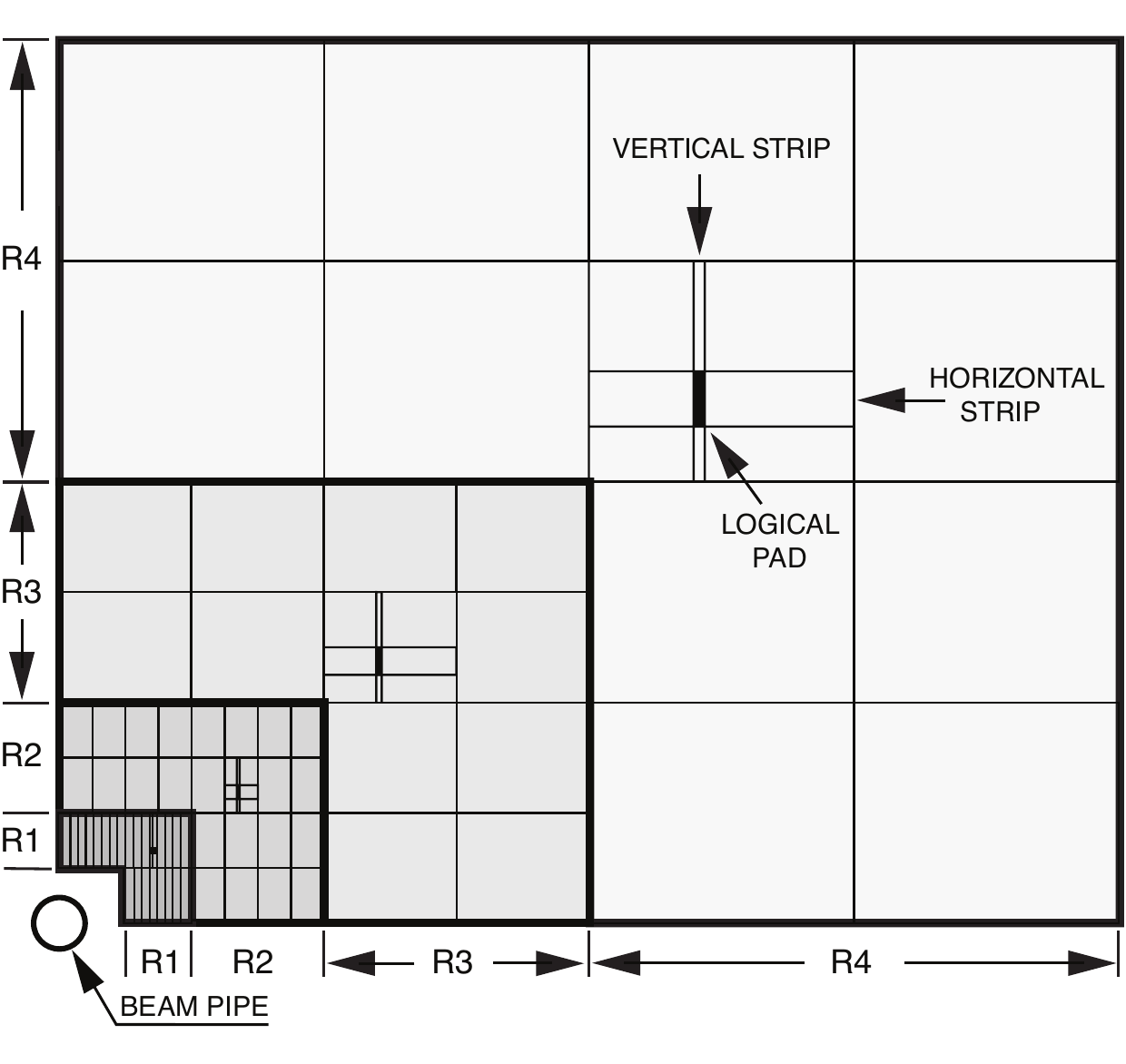}
    \caption{Front view of one quadrant of M2 showing the four regions. The intersection of a horizontal and a vertical strip defines a logical pad. The region and channel dimensions scale by a factor of two from one region to the following.}
    \label{fig:muon_chamber}
\end{figure}

Each muon station is subdivided into four regions, as shown in \Figref{fig:muon_chamber}, with different read-out schemes defining the $x,y$ resolutions. The dimensions of the logical pads were chosen such that their contribution to the \pt resolution was approximately equal to the multiple scattering contribution~\cite{LHCb-TDR-004}. 
These logical pads were obtained from the crossing of horizontal and vertical strips (either cathodic pads or group of wires), 
with the exception of the full M1 station and the innermost regions of stations M4 and M5, where the logical pads corresponded to physical channels on the detector and were readout directly. 

\begin{figure}[tb!]
    \centering
    \includegraphics[width=0.4\textwidth]{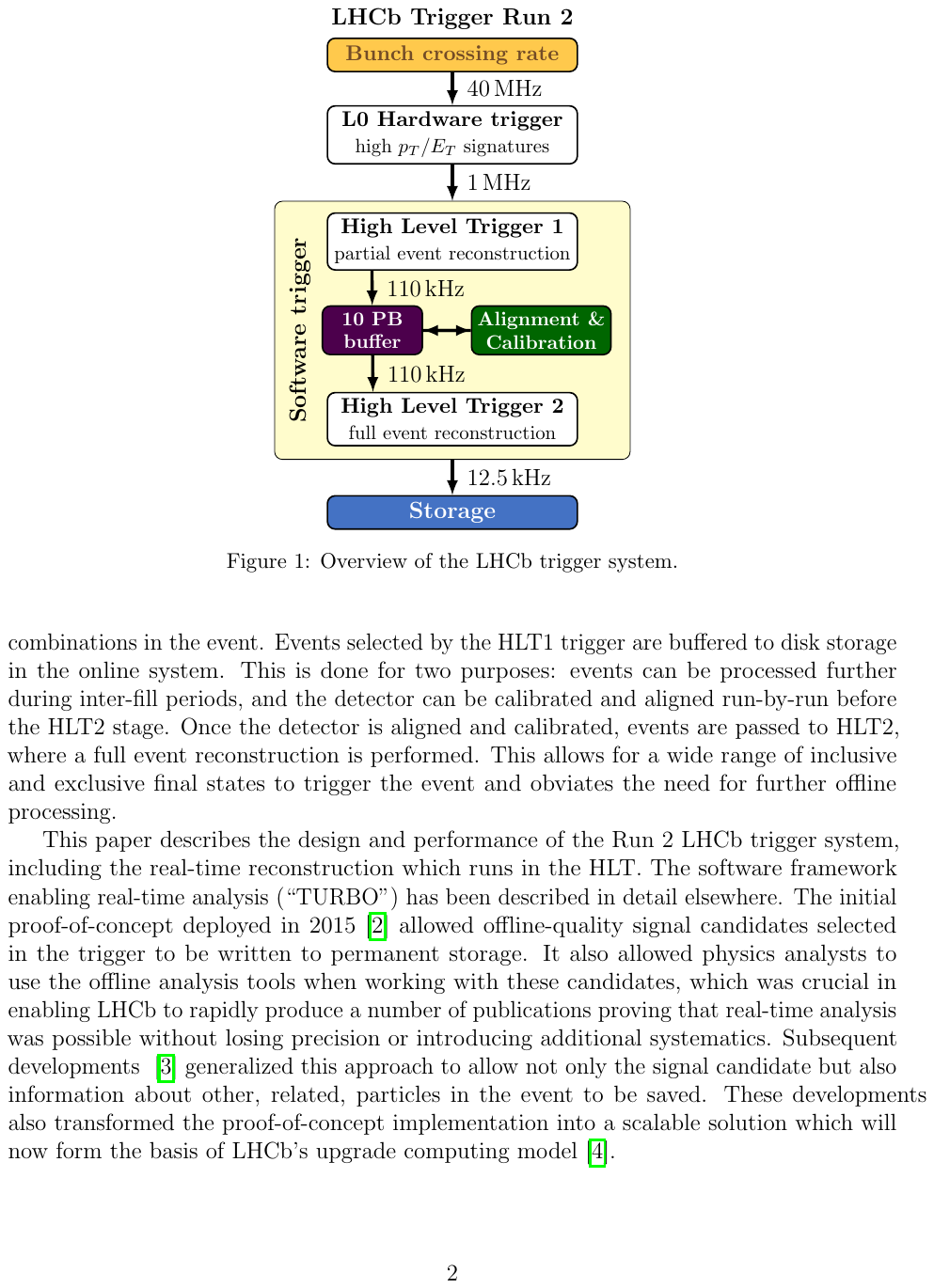}
    \caption{The LHCb trigger scheme in Run 2~\cite{Aaij_2019}.}
    \label{fig:trigger}
\end{figure}

A schematic diagram showing the trigger data flow in Run 2 is depicted in \Figref{fig:trigger}. 
The trigger and reconstruction scheme followed three basic steps:
\begin{itemize}
\item  A hardware trigger (\lone), based on selected calorimeter and muon information, to reduce the interaction rate of 20\mhz\footnote{Out of the total LHC bunch crossing rate of 40\mhz, there are about 30\mhz of inelastic collisions, of which around 2/3 are visible in the LHCb detector.}
to 1\mhz, which corresponded to the readout bandwidth of the detector. The \lone muon trigger was based on the coincidence of one hit in each of the five stations, selected in a projective Field Of Interest (FOI) defined in the $x-y$ plane, from which a muon standalone
  \pt reconstruction was performed with $\sim 20\%$ resolution~\cite{LHCb-TDR-004}. Candidate tracks above a \pt threshold of about $1.5\gevc$ were then used to build single and dimuon topologies.
\item  A first software stage (\hltone) based on partial reconstruction of tracks from the spectrometer, which allowed to put more strict constraints on the candidate \pt and IP. Concerning muons, the candidates from \lone were not used. Instead, a loose and efficient selection was performed, called \texttt{IsMuon}, based on the coincidence of hits in M2 to M5 stations, and combined with the information of the spectrometer. 
The muon hits were selected in a FOI
centered around the track extrapolation position on the muon stations: the number of hits required was two, three or four in the momentum ranges $3-6$, 
$6-10$ and above $10\gevc$, respectively, as expected from the muon penetration power in the iron absorbers~\cite{LHCb-PUB-2009-013}.
\item A more refined software trigger (\hlttwo), exploiting the full reconstruction of the detector information to reconstruct more complex signal topologies. Concerning muons, a better discrimination than \texttt{IsMuon} was achieved by using a likelihood variable (\texttt{MuonDLL}), built upon the uncorrelated sum of the spatial residuals 
of the muon hits with respect to the track extrapolation position in each station~\cite{LHCb-PUB-2009-013}, defined as: 
\begin{equation}
\label{eq:d2}
D^2 = \frac{1}{N}\sum_{i=1}^N \left[ \left( \frac{x^i_{\text{closest}} - x^i_{\text{track}}}{\text{pad}^i_x} \right)^2 +
\left( \frac{y^i_{\text{closest}} - y^i_{\text{track}}}{\text{pad}^i_y} \right)^2 \right],
\end{equation}
where the index $i$ runs over the $N$ stations containing hits inside the FOI, and the \textit{closest} coordinates represent the position of the hit which is closest to the
track extrapolation point. The hit residuals were  normalised to the logical pad size in the $x$ and $y$ directions, pad$_x$ and pad$_y$ respectively. 
The $D^2$ distribution for muons exhibits a narrow peak at 0, while hadrons satisfying the \texttt{IsMuon} criterion have a broader distribution.
Using the $D^2$ spectra of muons as a signal proxy and of protons as a background proxy (pions are instead contaminated by real muons from decays in flight),
the \texttt{MuonDLL} likelihood was defined, which measured the difference in probability for a candidate track to match the signal or background hypotheses.  
Using the above variable, on top of \texttt{IsMuon}, the misidentification probability for protons was kept at the 2-3 per mille level on the full momentum spectrum, with a muon efficiency above 90\%. For pions, similar misidentification probabilities were obtained only for momenta higher than 50\gevc, with decays in flight 
contributing for another few per mille at low momenta~\cite{Gandelman:1093941,LHCb-DP-2013-001}.
\end{itemize}

The \lhcb detector will be upgraded for Run 3 to sustain a factor of five increase in the instantaneous luminosity, up to $2\times 10^{33}$\,cm$^{-2}$s$^{-1}$. The 1\mhz readout limitation of the current detector will be removed, allowing for the full event rate to be processed in software without the need for a hardware stage~\cite{LHCb-TDR-012}.
For this reason, a full software trigger has been implemented, which will allow to select signal events with higher efficiency, 
and with the goal of achieving an order of magnitude increase in the physics bandwidth with respect to Run 2. 

In preparation of Run 3, the M1 station has been removed due to the much higher occupancy which will be reached in front of the calorimeter, where the station is located (\Figref{fig:muon_system}).
In addition, its main contribution, consisting in the improvement of the standalone muon \pt determination in the hardware \lone trigger is no longer relevant. When working on the implementation of the future software muon trigger lines, two aspects have to be taken into account:
\begin{itemize}
\item The need to keep a high efficiency at \hltone, with a smooth dependence on the running conditions and on the phase space, and with a fast execution time. Concerning the bandwidth, a high rejection power must be guaranteed against combinatorial background, especially important at low momentum. This background originates from tracks extrapolated to the muon detector and paired to accidental hits in the muon chambers due for example to other muons in the event or to electronic noise.
Accidental hits are expected to increase almost linearly with the luminosity.
\item The possibility to tune highly selective cuts in order to achieve very low mis-identification levels at \hlttwo, 
especially useful for example in rare decay searches. The full information from the \lhcb particle identification detectors may conveniently be used in this case as the constraints on the execution time are less stringent.
\end{itemize}

These goals can be achieved by following different approaches. In this paper we discuss a baseline strategy for the \hltone which is
an evolution of the present scheme.
This assumes that tracks in the spectrometer are reconstructed upfront, and that the muon identification is applied in two steps: a first step based on \texttt{IsMuon} as it is now,
plus a second step based on a correlated $\chi^2$ variable (Sec.~\ref{sec:newchi2}), which represents an improvement of the \texttt{MuonDLL}. At the \hlttwo stage,
the muon identification performance is further refined
by means of a multivariate classifier (Sec.~\ref{sec:MVAs}).

\section{Correlated \texorpdfstring{$\chi^2$}{chi2}}
\label{sec:newchi2}

The misidentification of charged pions and kaons to muons has an almost irreducible component due to decays in flight, 
together with a combinatorial component that is relevant especially at $p<10\gevc$, where the hit coincidence is less stringent.
The present muon identification algorithm was optimised
for a low occupancy scenario, 
without prioritising the rejection of the combinatorial background.
The higher luminosity of Run 3 will require instead to suppress this background more effectively, especially in the central detector regions where the occupancies are higher.

\begin{figure}[tb!]
\centering
\includegraphics[width=0.99\textwidth]{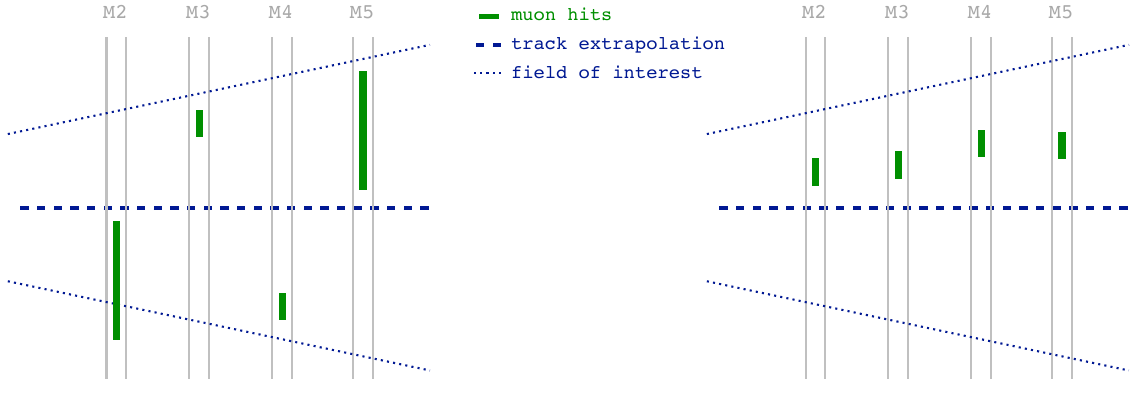}
\caption{Two different combination of muon hits having a similar value of \texttt{MuonDLL}: a combinatorial background event (left) and a clear muon pattern (right).}
\label{fig:muonDLL}
\end{figure}

An obvious limitation of the present approach based on the $D^2$ variable is that it does not include
the information from the multiple scattering experienced by charged particles traversing the calorimeter and the iron absorbers, as well as the correlation between the hit positions across the muon system.
The importance of taking into account these correlations
is evident in \Figref{fig:muonDLL}, where two very different hit combinations are shown, yet giving a similar \texttt{MuonDLL} value.
On the left, a random combination of hits scattered around the track extrapolation is shown, receiving contributions from uncrossed logical pads,
indicated by the larger error bars. 
Such events are more affected by electronic noise and spillover hits. 
On the right-hand side, a clear pattern of hits is visible, which are displaced with respect to the extrapolated track due to multiple scattering.

These two topologies can be discriminated by using 
a $\chi^2$ variable,  expressed as

\begin{equation}\label{eq:chi2corr}
\chicor = \delta\overrightarrow{x}^T \text{V}_x^{-1} \delta\overrightarrow{x} + \delta\overrightarrow{y}^T \text{V}_y^{-1} \delta\overrightarrow{y},
\end{equation}
where $\delta\overrightarrow{x}$ and $\delta\overrightarrow{y}$ are the distances, in the $x$ and $y$ directions, between the track extrapolation points and the closest hit positions, with indices
running over the stations  M2 to M5. The covariance matrices $\text{V}_x$ and $\text{V}_y$ both have a diagonal contribution from the detector resolution and off-diagonal terms quantifying the correlations induced by the multiple scattering. The diagonal terms corresponding to the hit position resolutions are of the form
\begin{equation}\label{eq:Var_RES}
  \text{V}^{\text{RES}}_{jj} = \left( \text{pad}^j_{x,y}/\sqrt{12} 
  \right)^2,
\end{equation}
where the pad size along $x$ and $y$  corresponding to the  muon hit in the given station are used.
The multiple scattering (MS) contributions have been modelled as
\begin{equation}\label{eq:Var_MS}
  \text{V}^{\text{MS}}_{jk} = \sum_{z_i < z_j,z_k} (z_j - z_i)(z_k - z_i) \sigma_{\text{MS},i}^2,
\end{equation}  
 where $z_{j,k}$ represent the coordinates of stations M2 to M5  along the beam axis, $z_i$
represents the coordinates of the main absorbers,  namely the calorimeters and the muon iron filters, as listed in \Tabref{tab:z/X0},
and $\sigma_{\text{MS},i}$
represents
the MS deviation. This term takes the expression~\cite{PDG2017}
\begin{equation}
\label{eq:ms}
\sigma_{\text{MS},i} = \frac{13.6~\text{MeV}}{\beta c p} \sqrt{\Delta z_i/X_0},
\end{equation}
where $p$ and $\beta c$ are the momentum and the velocity of the incident particle, respectively,
and $\Delta z_i/X_0$ is the thickness of the absorber at the given position $z_i$ in units of radiation length, also listed in \Tabref{tab:z/X0}.
The number of degrees of freedom (ndof), {\it i.e.} the order of the covariance matrices, corresponds to the number of stations in which a hit is found.
\begin{table}[htb!]
\centering
\begin{tabular}{c|cc}
\toprule
Absorber & $z$ position (m) & $\Delta z_i$/X$_0$ \\
\midrule
\ecal        & 12.8 & 25   \\
\hcal        & 14.3 & 53   \\
Muon filter 1  & 15.8 & 47.5 \\
Muon filter 2  & 17.1 & 47.5 \\
Muon filter 3  & 18.3 & 47.5 \\
\bottomrule
\end{tabular}
\caption{Position along the beam axis and thickness in units of radiation length for the main scattering media contributing to the multiple scattering experienced by particles reaching the muon detector.}
\label{tab:z/X0}
\end{table}

The probability for a muon to penetrate the iron absorbers and reach a given muon station depends on its momentum. In particular, below $6\,\gevc$ 
the probability to reach M4 and M5 stations can be substantially smaller than one, so that hits falling in the FOI of the track are in this case often due to accidental background. For this reason, in that momentum interval  
only the hits on M2 and M3 stations
are included in the \chicor computation.

\begin{figure}[tb!]
  \centering
  \includegraphics[width=0.47\textwidth]{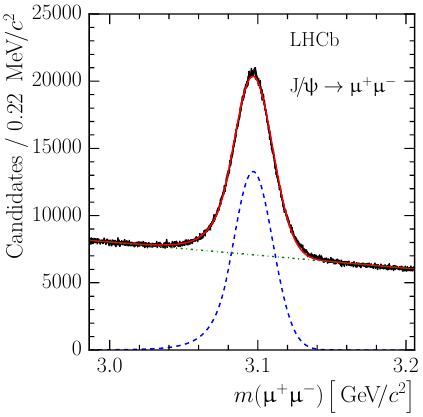}
  \includegraphics[width=0.47\textwidth]{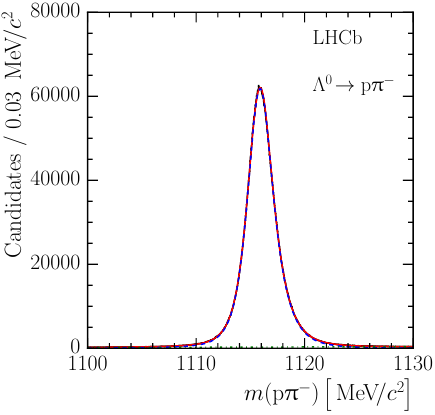}
    \caption{Typical invariant mass distributions for $J/\psi\to\mu^+\mu^-$ (left) and $\Lambda\to p\pi^-$ (right) calibration samples.
    The superimposed fit (red line) is composed of a signal (dashed blue) and background (dotted dash green) component~\cite{Lupton:2134057}.}
  \label{fig:calib_samples}
\end{figure}

The performance of the \chicor variable is evaluated on muons and protons from data control samples collected in 2016. Data samples are preferred over simulation as the performance of the algorithms is very sensitive to the occupancy of the detector. The number of hits in the detector is dominated by low energy background from particles with energies below the simulation thresholds. Hence, data calibration samples are a better proxy for the background expected in Run 3. 
An abundant source of muons is provided by $J/\psi \to \mu^+\mu^-$ decays: by requiring the reconstructed $J/\psi$ to have a large flight distance significance and good decay vertex quality, most of the combinatorial background from the tracks originating from the primary vertex is removed, and the sample gets enriched by $B \rightarrow J/\psi X$ candidates. To further reduce the background, one of the decay tracks, the $tag$ muon, is required to be positively identified in the muon detector; the other track, the $probe$ muon, is unbiased with respect to particle identification and trigger requirements and it is therefore used to measure the algorithm performances.
Protons are selected from $\Lambda \to p \pi^-$ decays with vertex quality criteria and detachment of the decay vertex from the primary one. In addition, the invariant mass obtained by assigning the $\pi$ mass to the two daughters is required to be outside the nominal $K^0_S$ mass window. Examples of mass spectra for muon and proton calibration samples are shown in \Figref{fig:calib_samples}.

For both samples, the residual background contribution is subtracted by using the \sPlot method~\cite{Pivk:2004ty}.
To perform unbiased studies, the muon and proton samples have been weighted in order to equalise their momentum, transverse momentum, and track multiplicity spectra.  
In addition, since the main challenge for Run 3 is the fivefold luminosity increase with respect to Run 2, a weighting procedure that adds more emphasis on high multiplicity events is applied to each calibration sample. 
Since there is not enough data to accurately emulate the upgrade conditions, the samples have been weighted in such a way to reproduce an occupancy spectrum which is in-between the two actual running conditions.
\begin{figure}[tb!]
\centering
\includegraphics[height=0.5\textwidth]{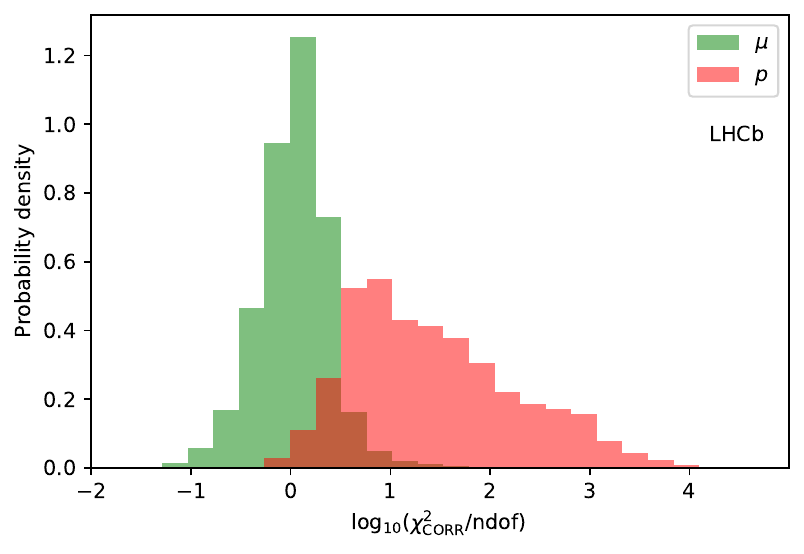}
\caption{Spectrum of the \chicor, normalised to the degrees of freedom, for muons and protons as evaluated on 2016 calibration samples.}
\label{fig:chi2cor}
\end{figure}

\begin{figure}[tb!]
  \raggedright
  \includegraphics[width=0.32\textwidth]{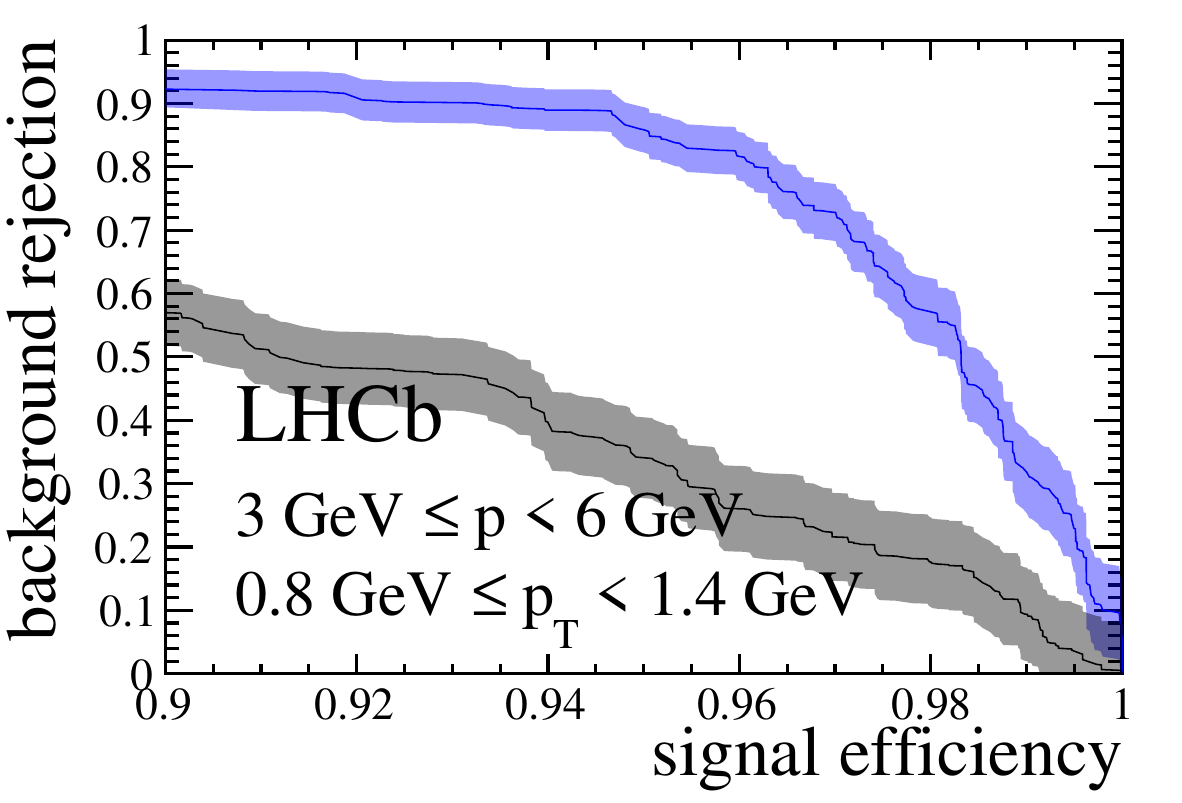}
  \includegraphics[width=0.32\textwidth]{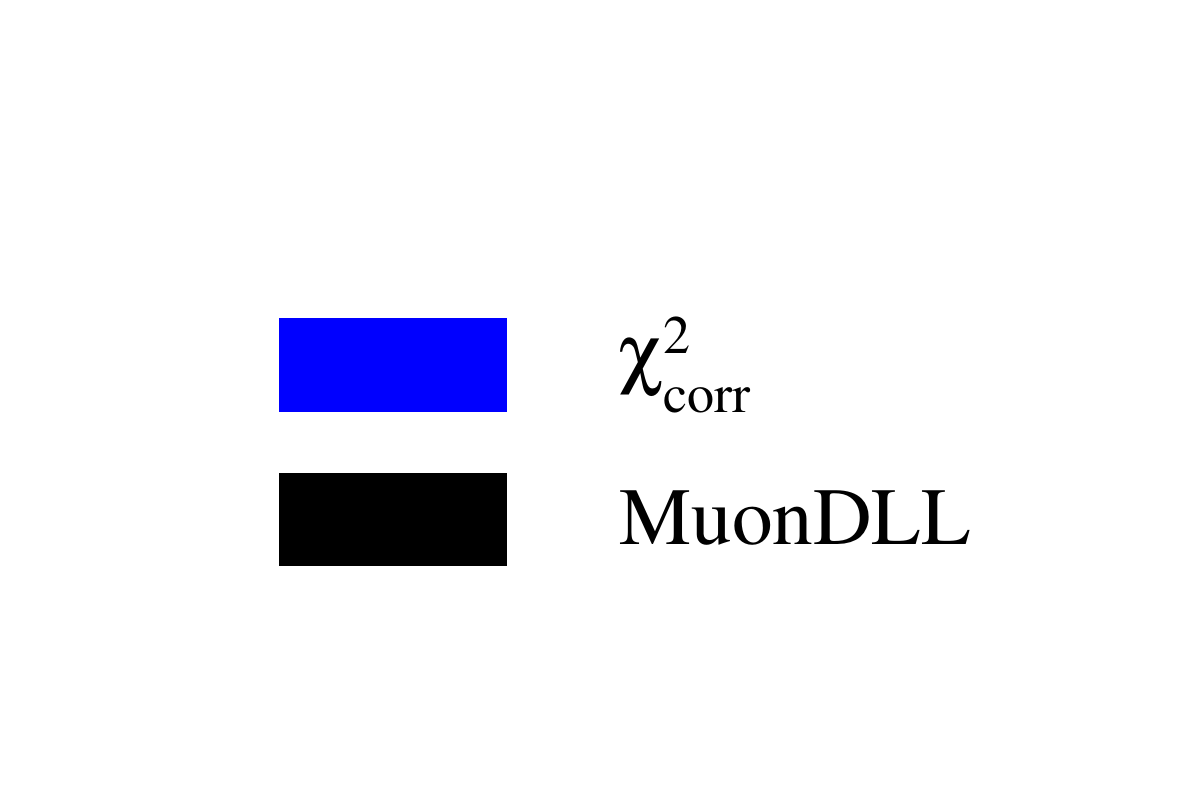}\\
  \includegraphics[width=0.32\textwidth]{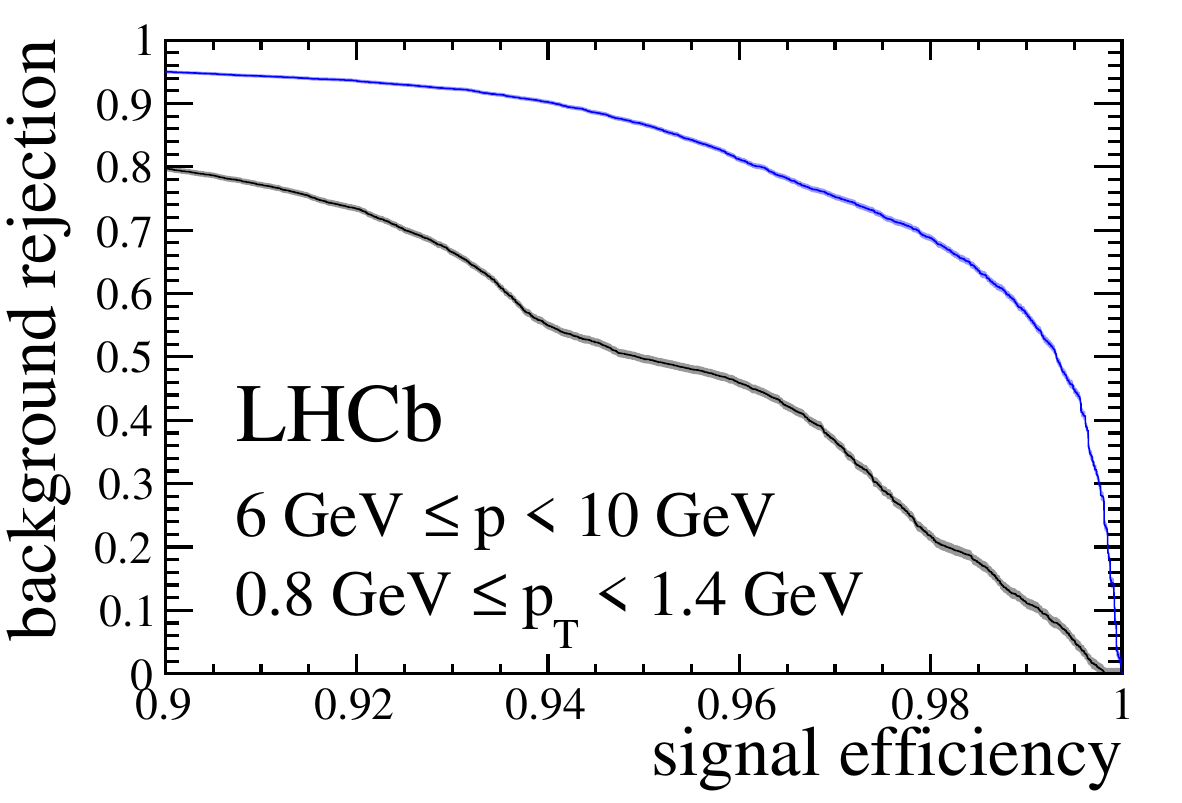}
  \includegraphics[width=0.32\textwidth]{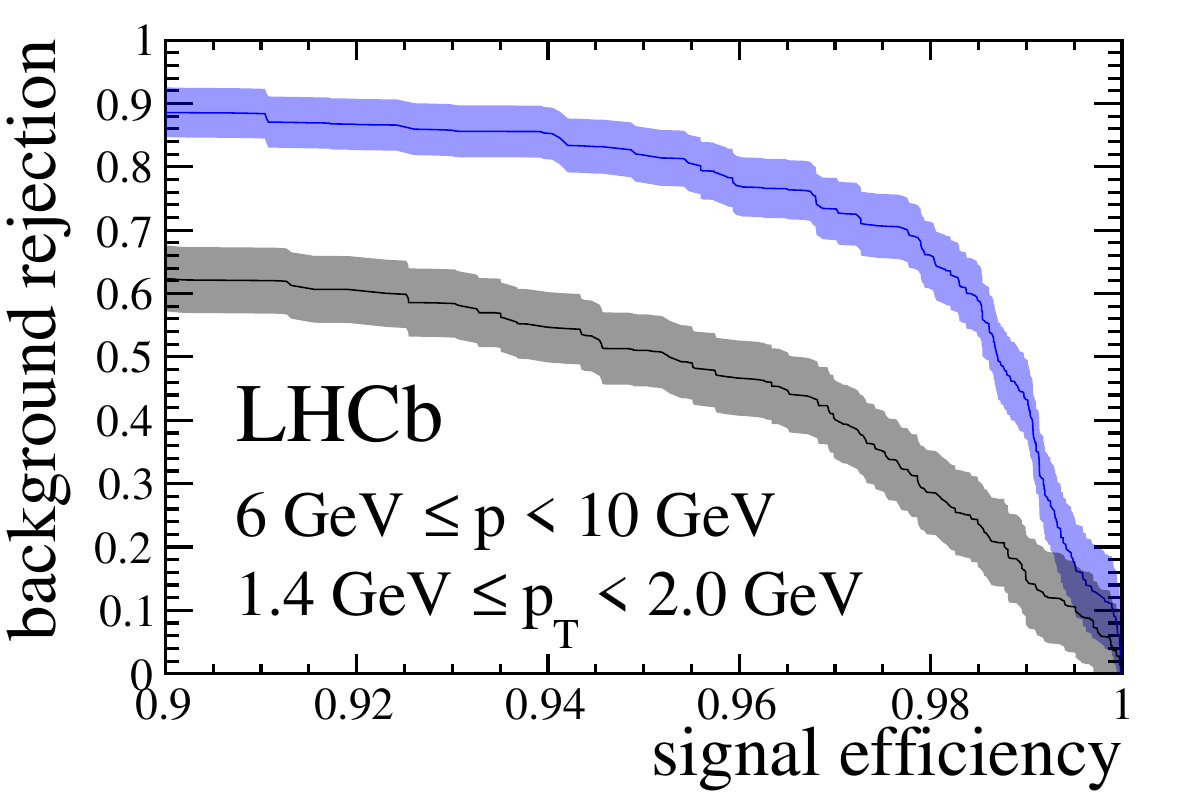}\\
  \includegraphics[width=0.32\textwidth]{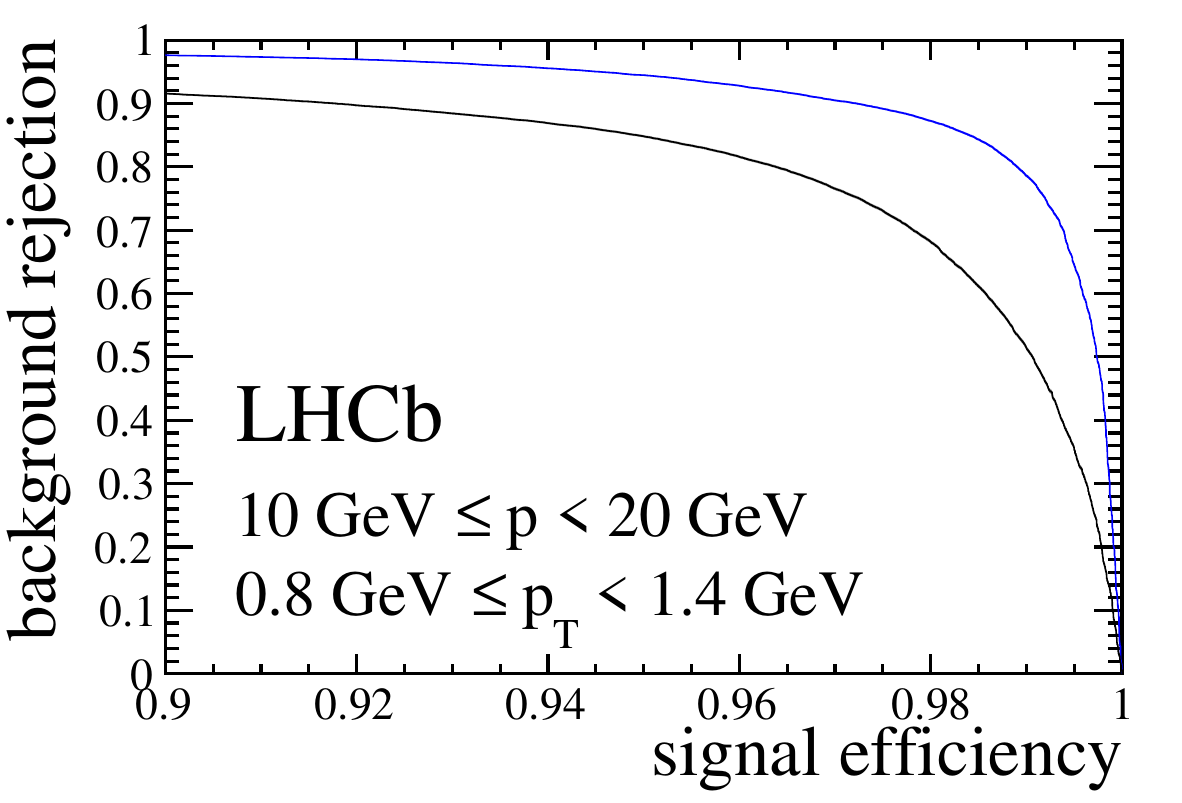}
  \includegraphics[width=0.32\textwidth]{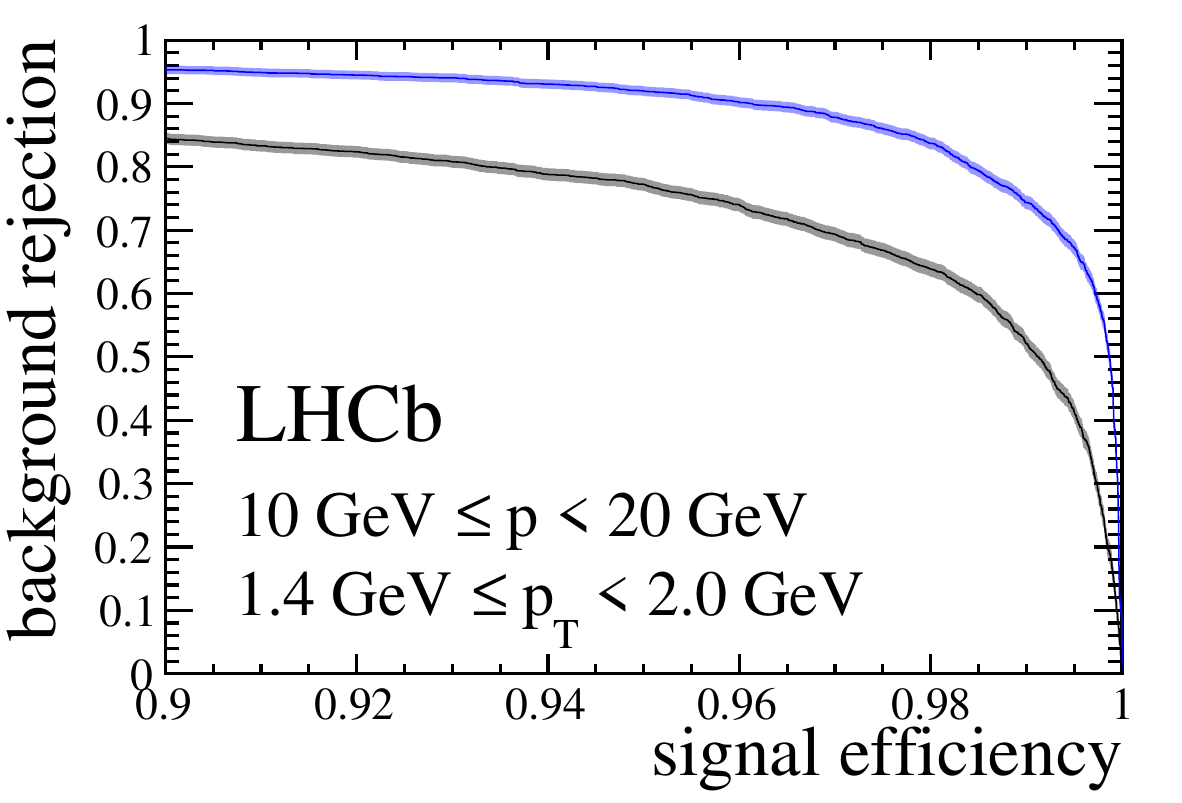}
  \includegraphics[width=0.32\textwidth]{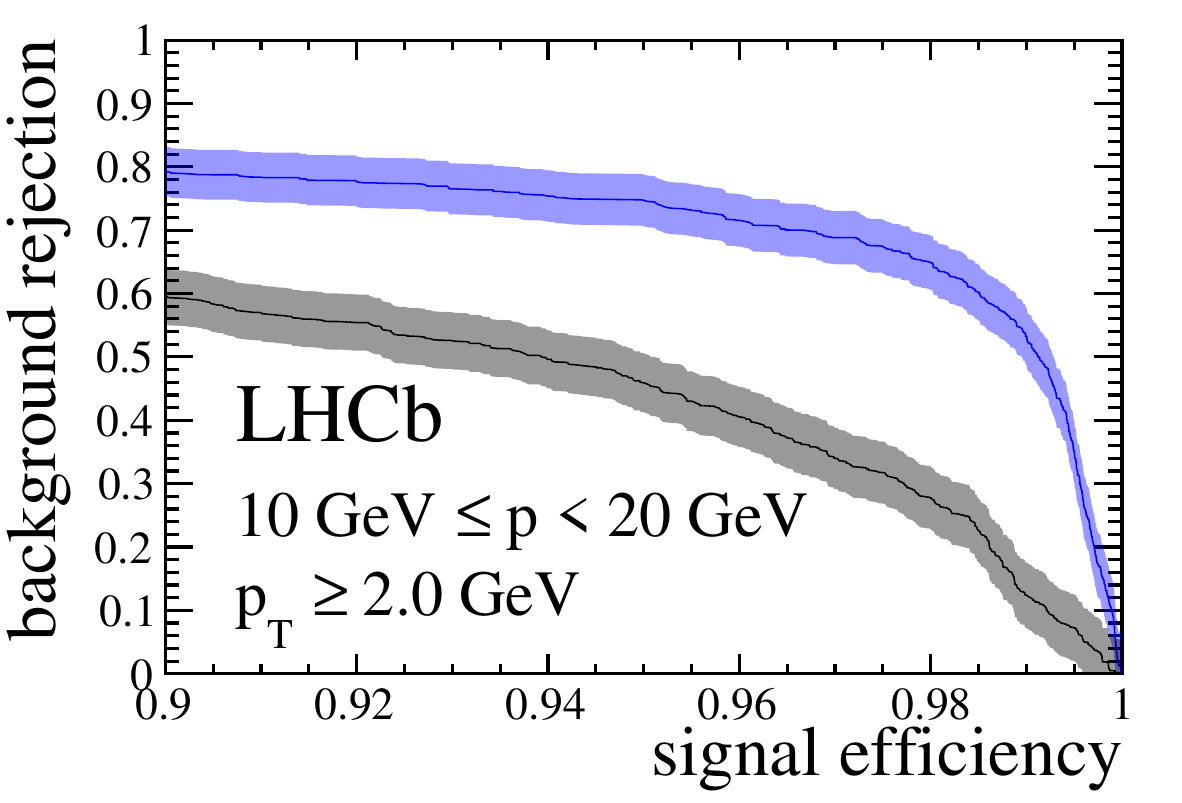}
  \includegraphics[width=0.32\textwidth]{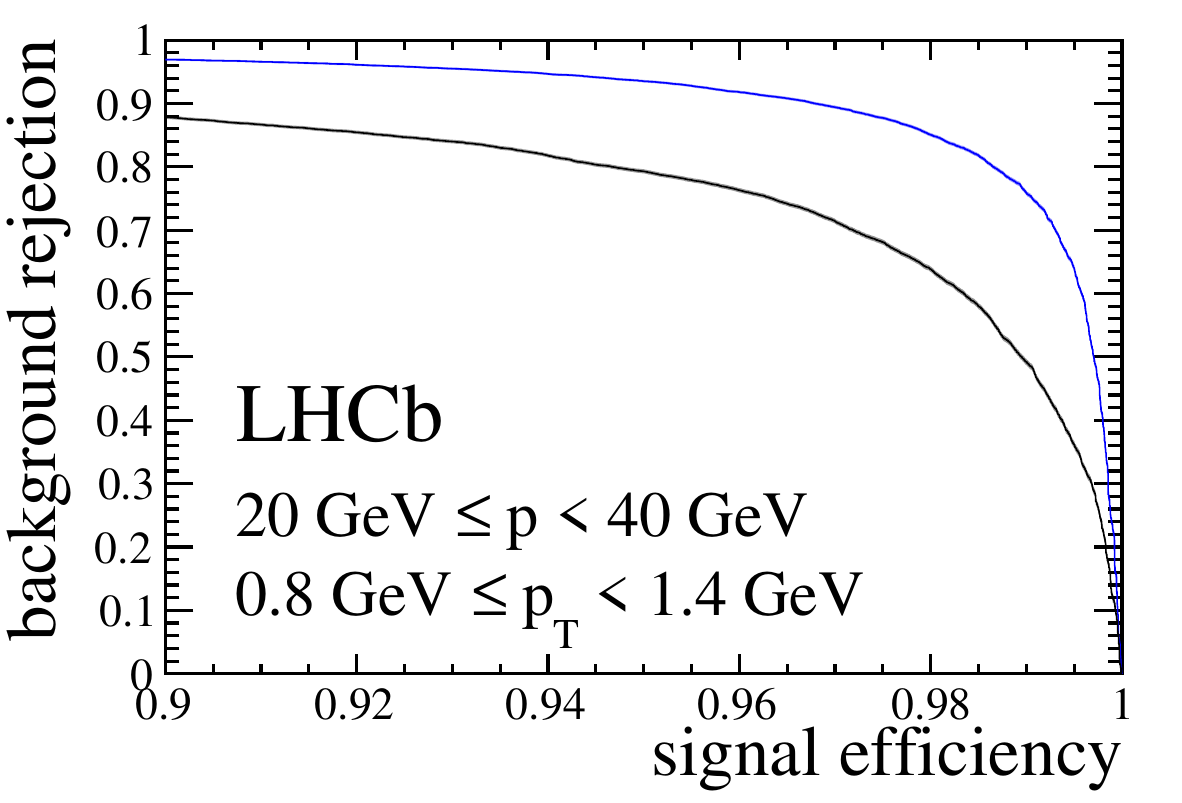}
  \includegraphics[width=0.32\textwidth]{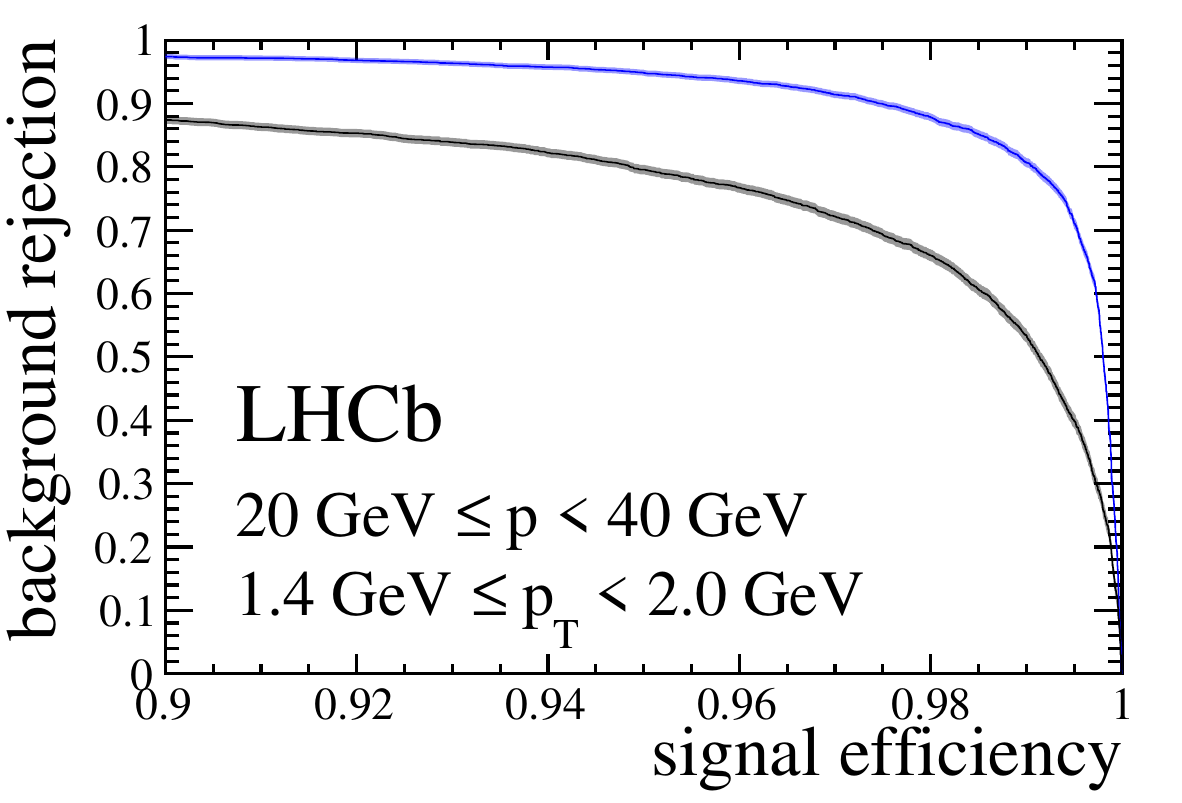}
  \includegraphics[width=0.32\textwidth]{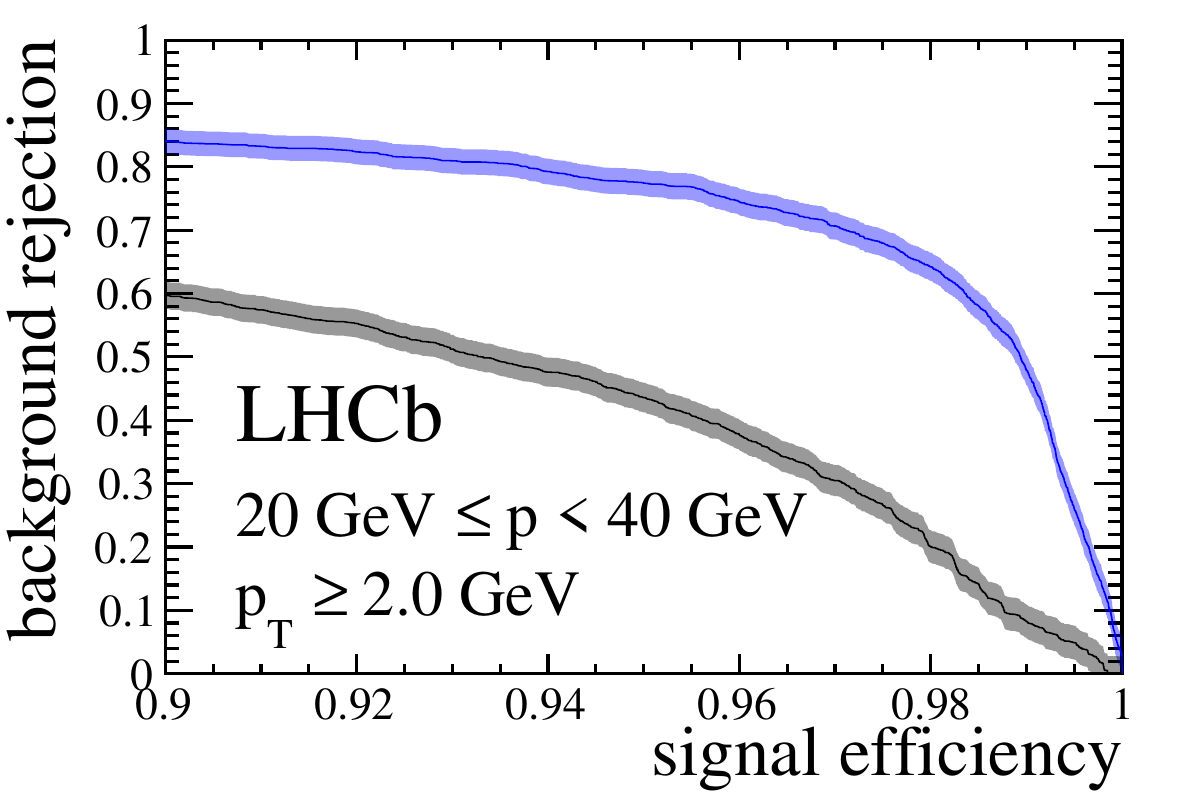}
  \includegraphics[width=0.32\textwidth]{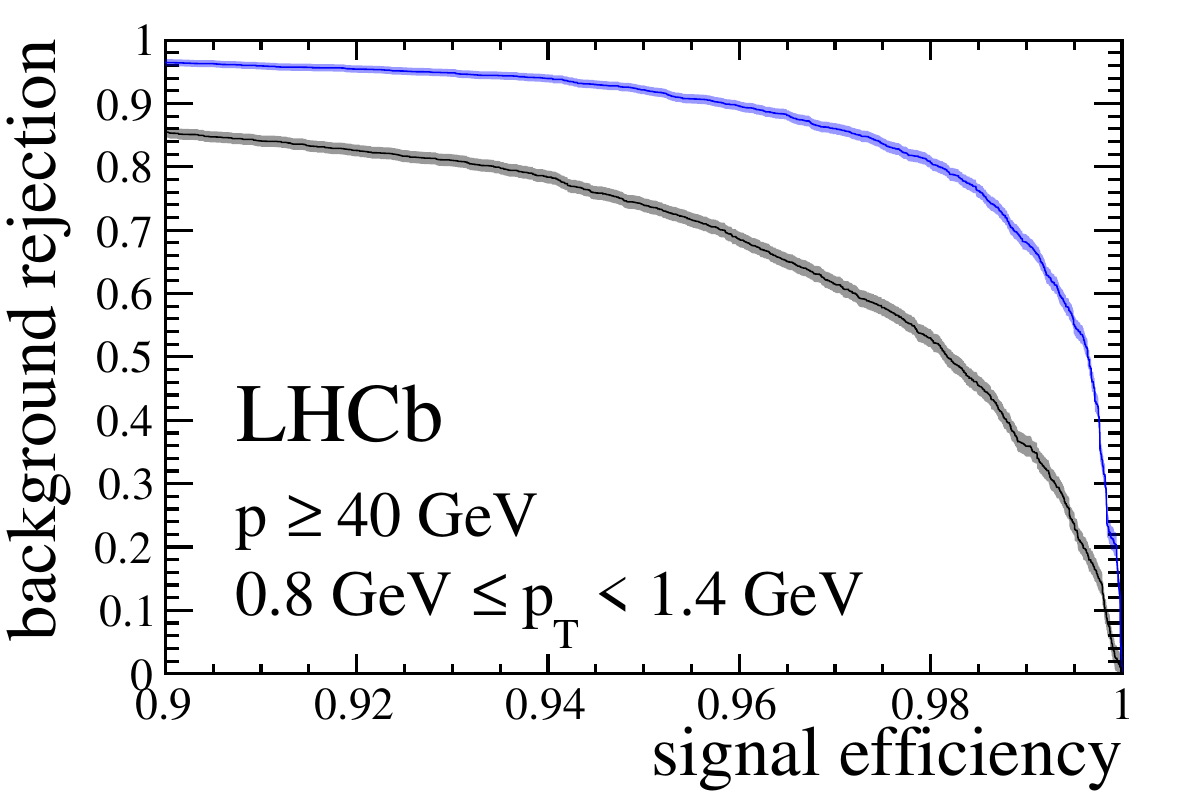}
  \includegraphics[width=0.32\textwidth]{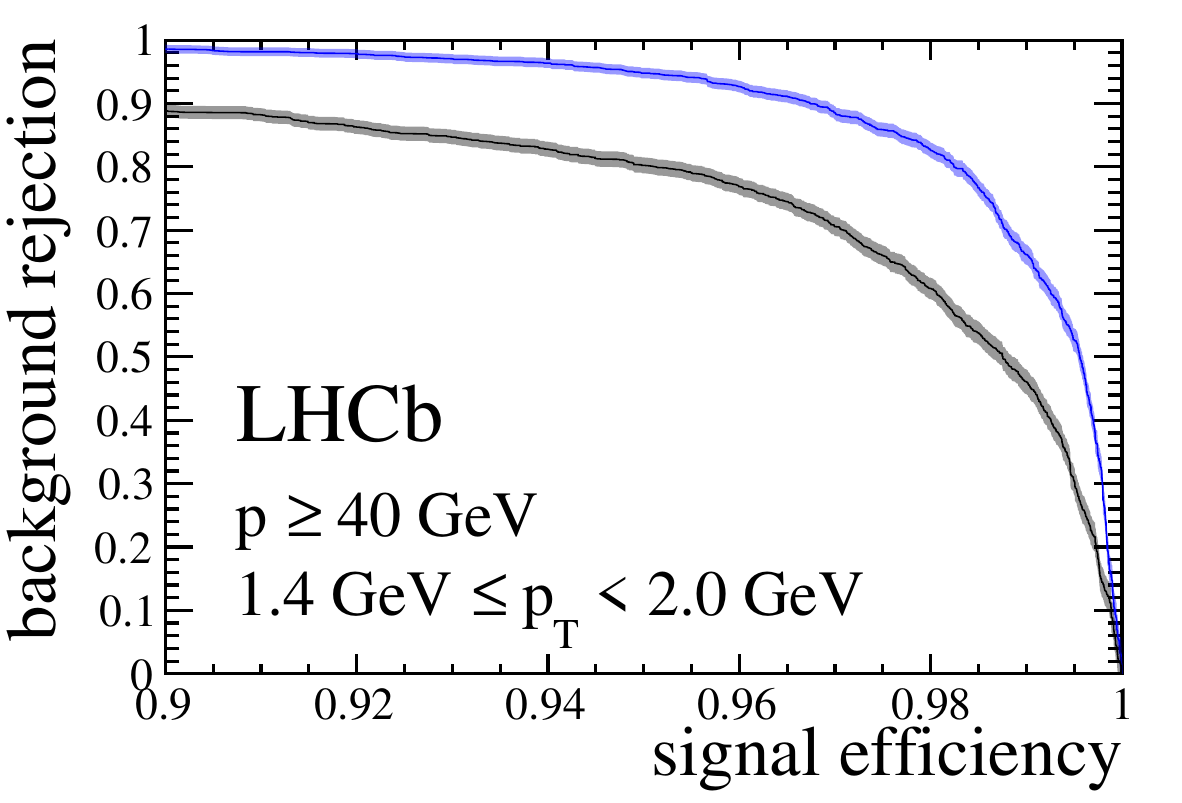}
  \includegraphics[width=0.32\textwidth]{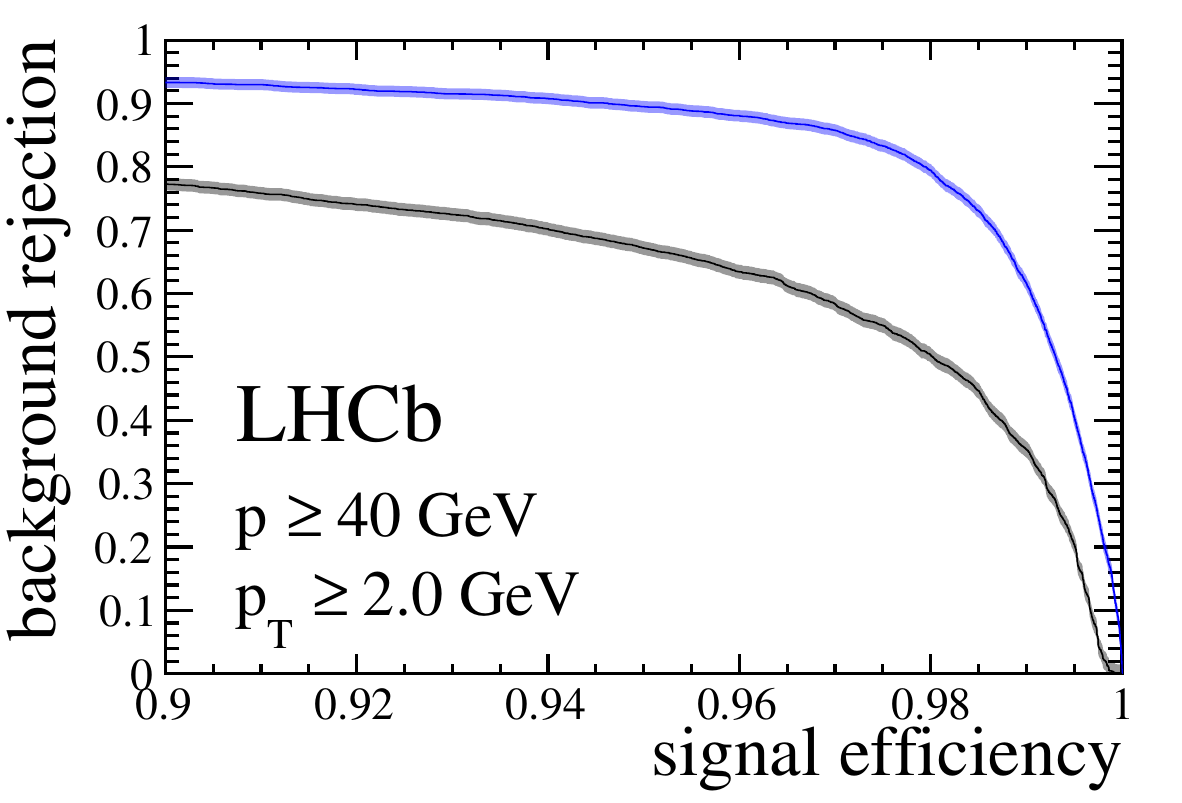}  
  \caption{Proton rejection as a function of muon efficiency for tracks satisfying \texttt{IsMuon} obtained with the \chicor (blue) and \texttt{MuonDLL} (black) variables on 2016 calibration data. Low momentum bins, which are not covered by the calibration samples, are not shown. The bands represent statistical uncertainties.}
  \label{fig:run_2_chi2_proton}
\end{figure}

As a result, in \Figref{fig:chi2cor} the \chicor spectrum for muons and protons satisfying the \texttt{IsMuon} requirement is shown, demonstrating a
good separation between signal and background.
A quantitative comparison between the performance of the \chicor and \texttt{MuonDLL} variables is shown in \Figref{fig:run_2_chi2_proton},
where the proton rejection as a function of the muon efficiency, ROC curve in the following, is displayed
for tracks satisfying the \texttt{IsMuon} requirement. The ROCs are
shown in different momentum and transverse momentum intervals, which allow to probe the response of the muon identification algorithms in different regions of the detector
and in different momentum regimes. 
The performance of the \chicor variable is definitely better than the \texttt{MuonDLL} in all regions of the phase space, and especially at low momenta.
In particular, at muon efficiency of $\sim 98\%$, which is a good working point for efficient trigger selections, the gain in background rejection is a factor $\sim 1.4$
in the region $p>10\,\gevc ,~\pt < 2 \, \gevc $, and exceeds  a factor of $2$ in the rest of the phase space.

\subsection{Performance in \hltone}
\label{ssec:perf}

As discussed in Sec.~\ref{sec:Introduction}, for the \hltone in Run 3 it will be crucial to guarantee a high efficiency for muons and a fast execution time of the algorithms. Moreover, a tighter rejection against combinatorial background with respect to the present \texttt{IsMuon} selection will be certainly needed.

Given the good performances of the \chicor variable in rejecting protons, which constitute pure combinatorial background to the muon detector, we consider as interesting to provide the
rejection estimates on trigger unbiased events, which are mostly populated by pions, selected from a Run 2 data sample without any trigger requirement.
As a preliminary selection for this benchmark, the events are filtered by requiring at least one track to satisfy \texttt{IsMuon} and the cuts $p_T>800\mevc$ and $\text{IP}\chi^2>35$~\footnote{The
impact parameter $\chi^2$, $\text{IP}\chi^2$, is defined as the difference in the primary vertex fit $\chi^2$ with and without the given track. Detached tracks, for example those coming from a $B$ decay, have larger $\text{IP}\chi^2$ values.},
which represent the main requirements of the Run 2 \hltone single muon line. The rejection is therefore computed relatively to the above selection, and thus represents the improvement with respect to the present \hltone, and by removing the \texttt{L0} trigger. To select high multiplicity events, only those having at least $3$ primary vertices (nPVs) are used, whereas average Run 2 events have one primary vertex.

This study is done in three momentum intervals, $3-6$, $6-10$ and $p>10\gevc$, since the number of hits selected by \texttt{IsMuon} is different in each one, as described in Sec.~\ref{sec:Introduction}.
In each interval, a \chicor cut with $\sim 98\%$ muon efficiency, as evaluated on muon calibration data, is chosen.
The results are shown in Tab.~\ref{tab:reduc} and demonstrate the effectiveness of this variable in rejecting about half of the trigger unbiased events, with a very small efficiency loss, on top of the Run 2 \hltone muon selection.
In particular, the highest rejection is achieved for $6 < p < 10\gevc$, where the fraction of pion decays in flight is lower with respect to $3<p<6\gevc$, and the multiple scattering correlations provide sensible discrimination as the momentum is not too high.

\begin{table}[htb!]
\centering
\begin{tabular}{c|c}
\toprule
Momentum range & Rejection factor \\
\midrule
    $3<p<6 \,\gevc$ & $ 1.8 $ \\ 
    %\hline 
    $6<p<10\,\gevc$ & $ 3.2 $ \\ 
    %\hline 
    $p>10 \, \gevc$ & $ 2.2$ \\ 
\bottomrule
\end{tabular}
\caption{Rejection factors of the \chicor variable on trigger unbiased events, for a muon efficiency of $\sim 98\%$. The rejection is evaluated on top of the $p_T>800\mevc$, $\text{IP}\chi^2>35$, \texttt{IsMuon} and nPVs~$\geq3$ requirements.}
\label{tab:reduc}
\end{table}

Finally, the \chicor execution time is tested within the \hltone upgrade sequence.
Throughput tests\footnote{On nodes mounting two Intel Xeon E5-2630 v4 CPUs at $2.20$ GHz (40 threads/node).}
are performed on simulated Run 3 data and show a \chicor resource usage of about $0.4\%$ out of a total \hltone throughput rate of $\sim36$ MHz, and in view of a data taking rate of $30$ MHz.
This result makes the \chicor well suited for a usage in the upgraded \hltone trigger of the experiment.

\section{Multivariate algorithm}
\label{sec:MVAs}
At the second stage of the trigger, the timing budget allows to use more complex algorithms.
Besides the spatial information, each muon hit also carries two different time counters, one for each view, $x$ and $y$. 
The number of views, {\it i.e.} the fact that the hit is crossed or uncrossed, also provides valuable
information, as noise or spillover hits typically have one view only. 
This information, along with its correlations, can be exploited in a multivariate operator.
To this purpose, a recent variant of gradient tree boosting available in
the CatBoost library from Yandex~\cite{prokhorenkova2018catboost} has been implemented~\cite{ml2019sweights-acat, mlsplot-jinst}. It uses oblivious decision trees as weak learners, as explained in the following.

\begin{figure}[ht!]
\begin{subfigure}{0.40\textwidth}
    \centering
    \includegraphics[width=\textwidth]{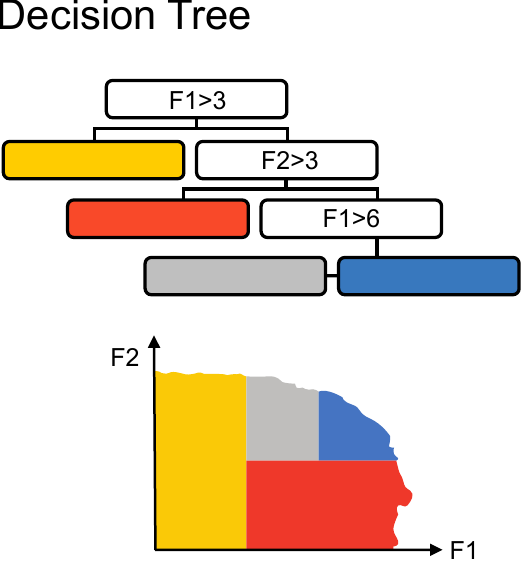}
\end{subfigure}
\hfill
\begin{subfigure}{0.45\textwidth}
    \centering
    \includegraphics[width=\textwidth]{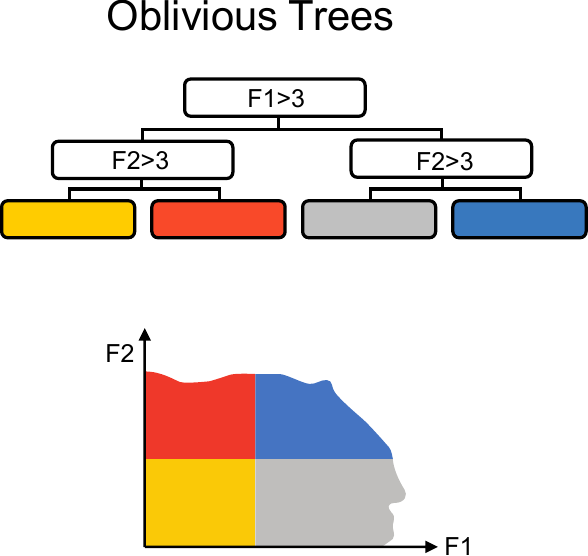}
\end{subfigure}
\caption{Classic versus oblivious decision trees. Reproduced from \cite{matrixnet}.}
\label{fig:oblivious-tree}
\end{figure}

A regular decision tree selects each split independently, while an oblivious decision tree has the same split on each level. The difference is illustrated in
Fig.~\ref{fig:oblivious-tree}. An oblivious decision tree is less expressive but is much faster to evaluate, as it makes possible to unwrap the tree into a
table and look up the correct leaf in one operation, instead of the multiple conditional jumps of a regular tree. According to a benchmark study by the CatBoost authors, which we were able to reproduce, this provides 30--100 faster prediction compared to the competing state-of-the-art gradient boosting libraries~\cite{catboostbenchmark}.

For the muon identification, five variables for each muon station M2 to M5 are used as input to the CatBoost algorithm:
\begin{itemize}
\item $x_{\text{res}}$: the difference between the closest hit $x$ position and the track extrapolation, normalised to the total uncertainty;
\item $y_{\text{res}}$: the difference between the closest hit $y$ position and the track extrapolation, normalised to the total uncertainty;
\item $t_x$: the time of the $x$ view;
\item $dt = t_x-t_y$: the temporal difference between the $x$ and $y$ views;
\item $N_{\text{views}}$: the number of views.
\end{itemize}
The uncertainty in the residuals $x_{\text{res}}$ and $y_{\text{res}}$ contains the pad size and the contribution from the multiple scattering (Eq.~\ref{eq:ms}), summed in quadrature. 
In addition to the hit information, for each event the track extrapolation $x$ and $y$ coordinates on M2 are used, to allow the algorithm to discriminate between different detector regions. Finally, the aforementioned \chicor variable (Sec.~\ref{sec:newchi2}) of the track is added.
This set of up to 23 variables per event has been found to be the smallest one containing the maximum associated information, without introducing excessive correlations. 
This feature is very important in order to decrease the complexity and hence the computation time of the operator in the trigger.

The classifier is trained using samples from 2016 data, to which the \texttt{IsMuon} requirement is applied. Since the \texttt{IsMuon} algorithm is very fast to execute and already rejects around 99\% of background, evaluating the classifier only for events that pass the \texttt{IsMuon} requirement allows to significantly reduce the computational cost and to focus on reducing the remaining background. The data samples used in the training are the same calibration samples used for the \chicor evaluation:
\begin{itemize}
    \item muons from $J/\psi \rightarrow \mu^+\mu^-$ decays,
    \item protons from $\Lambda \rightarrow p\pi^-$ decays,
    \item pions from $D^{*-} \rightarrow D^{0} (\rightarrow K^- \pi^+) \pi^- $ decays.
\end{itemize}
While protons represent pure combinatorial events, the pion sample is added to boost the training statistics and accounts for another source of classification error due to particles that decay in flight to muons before reaching the muon stations.
These samples have been treated as described in Sec.~\ref{sec:newchi2}, including kinematic reweighting, background subtraction and multiplicity weights.

To deal with negative sWeights, the solution proposed in Ref.~\cite{ ml2019sweights-acat} is used, which consists of first using a machine learning regression to estimate the expected sWeight in each point of the training variable phase space and, second, using the expected sWeight as event weight during classification.

\begin{figure}[htb!]
    \raggedright
    \includegraphics[width=0.32\textwidth]{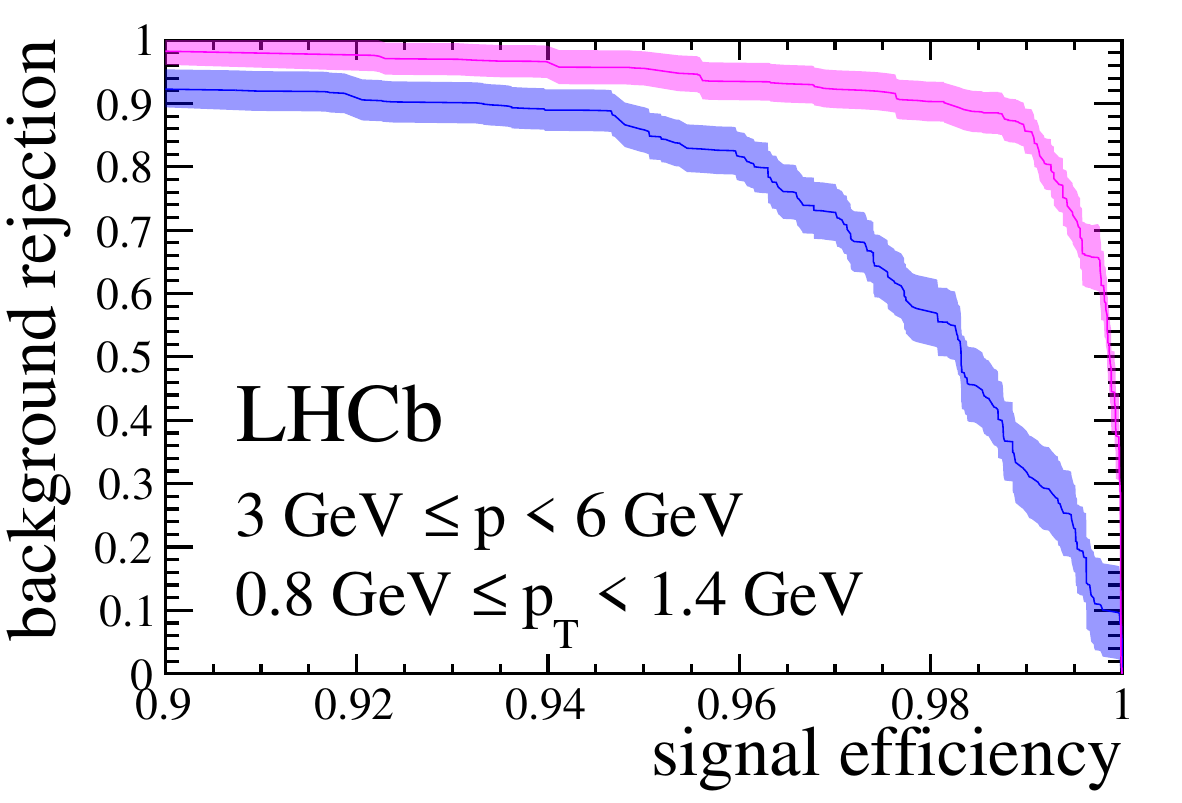}
    \includegraphics[width=0.32\textwidth]{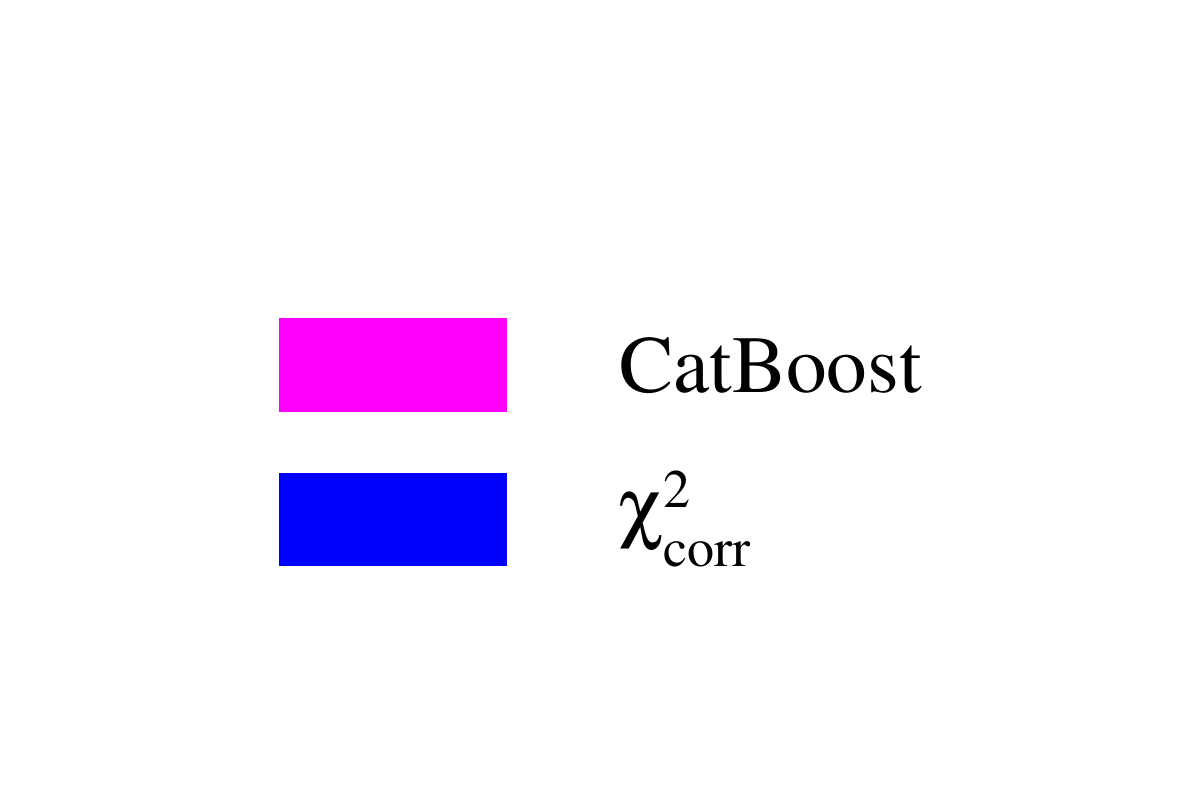}\\
    \includegraphics[width=0.32\textwidth]{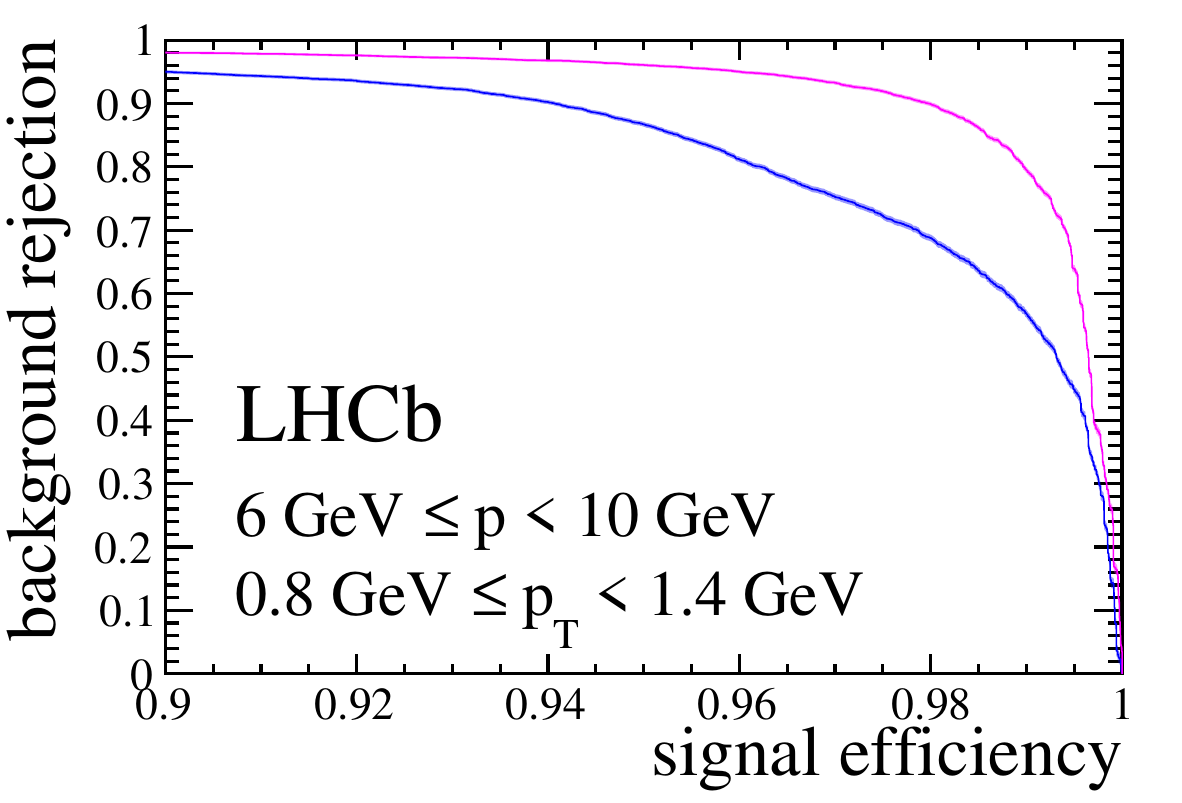}
    \includegraphics[width=0.32\textwidth]{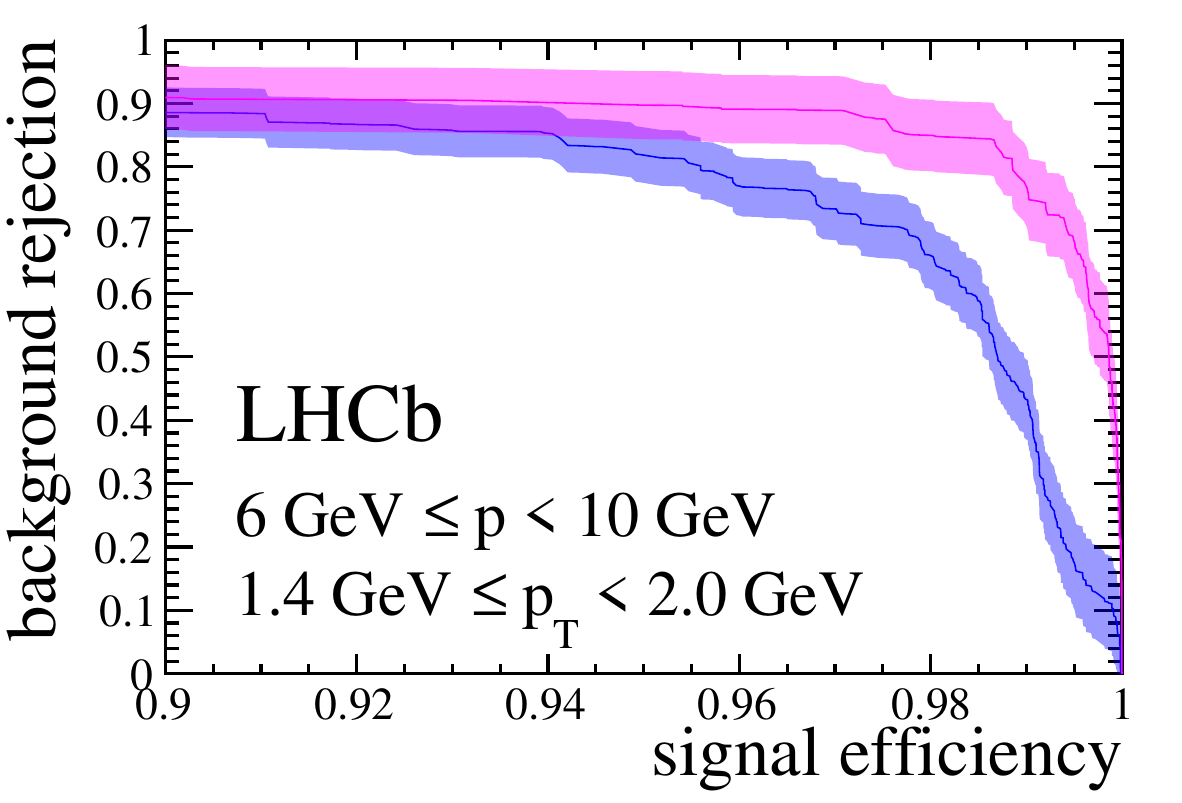}\\
    \includegraphics[width=0.32\textwidth]{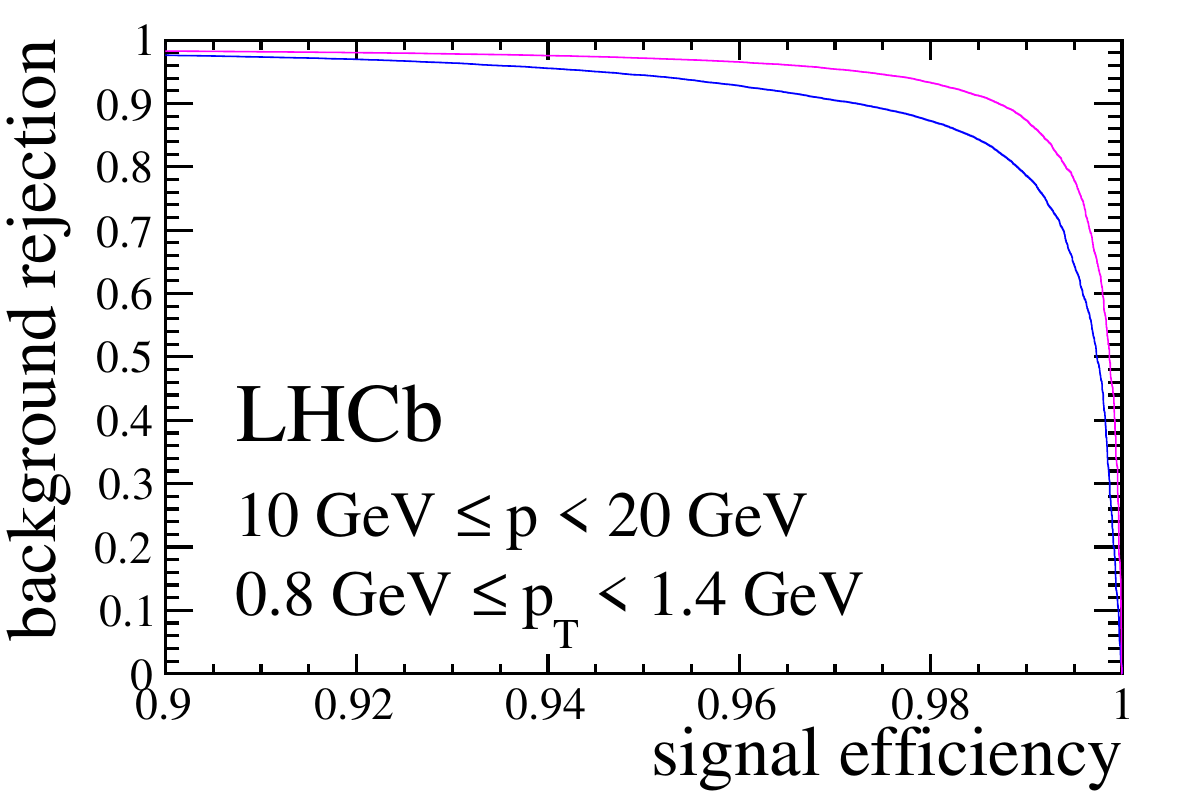}
    \includegraphics[width=0.32\textwidth]{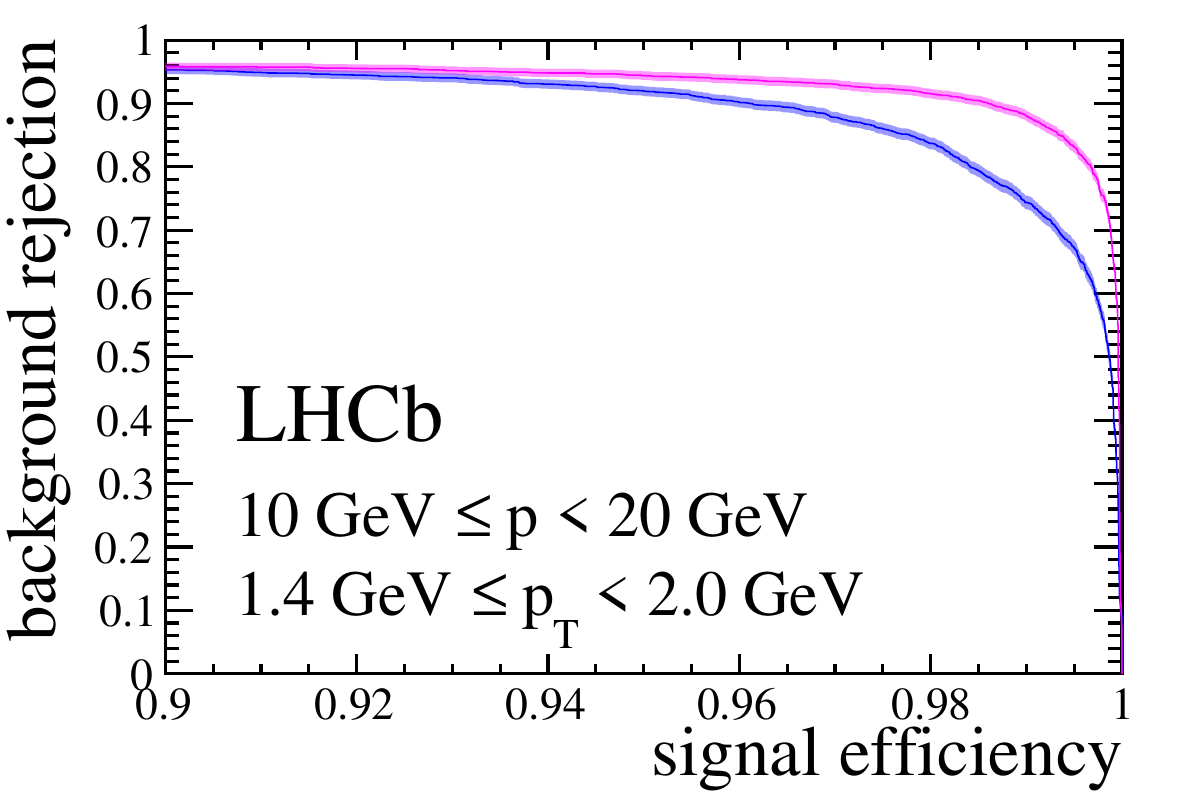}
    \includegraphics[width=0.32\textwidth]{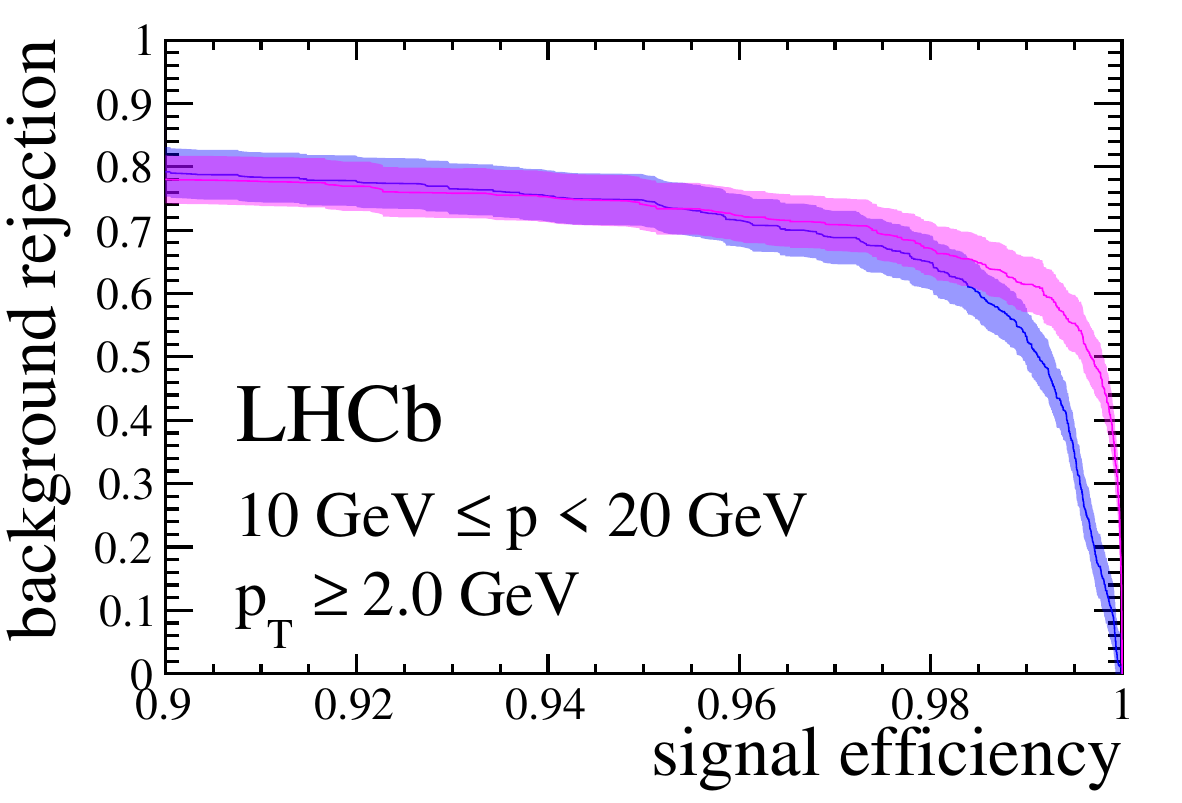}
    \includegraphics[width=0.32\textwidth]{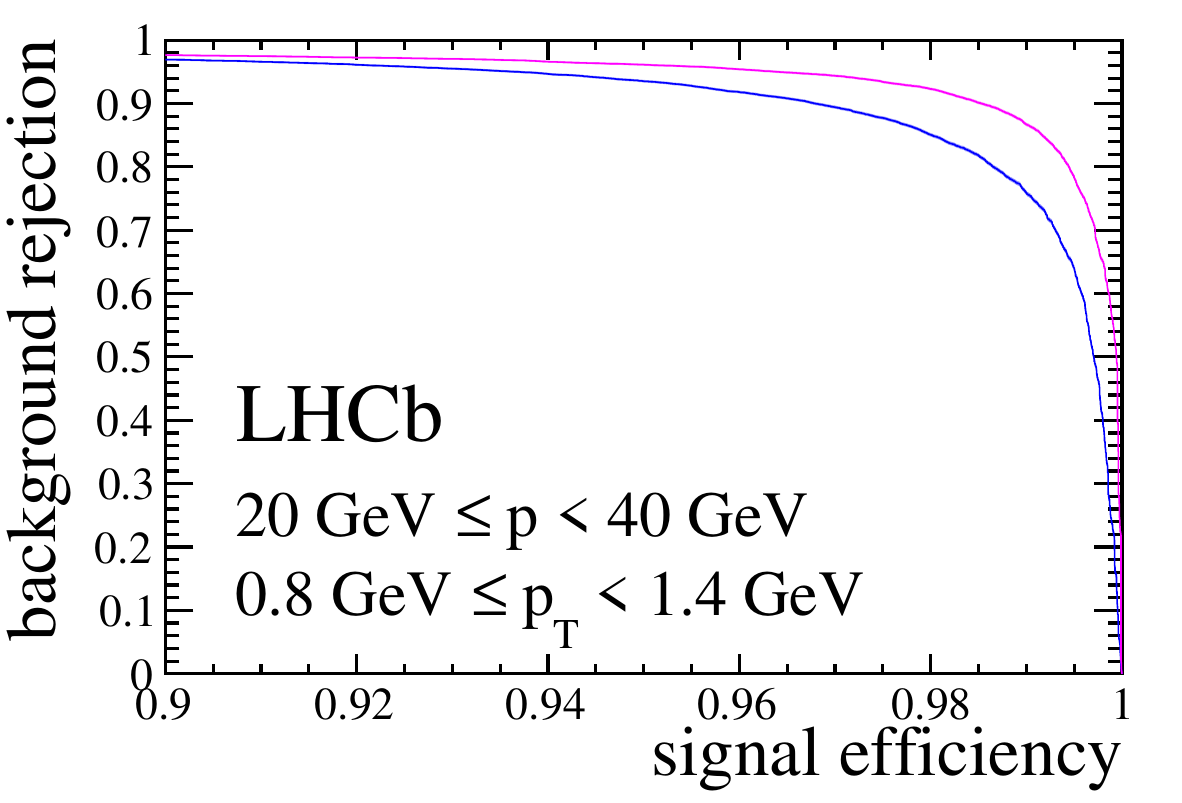}
    \includegraphics[width=0.32\textwidth]{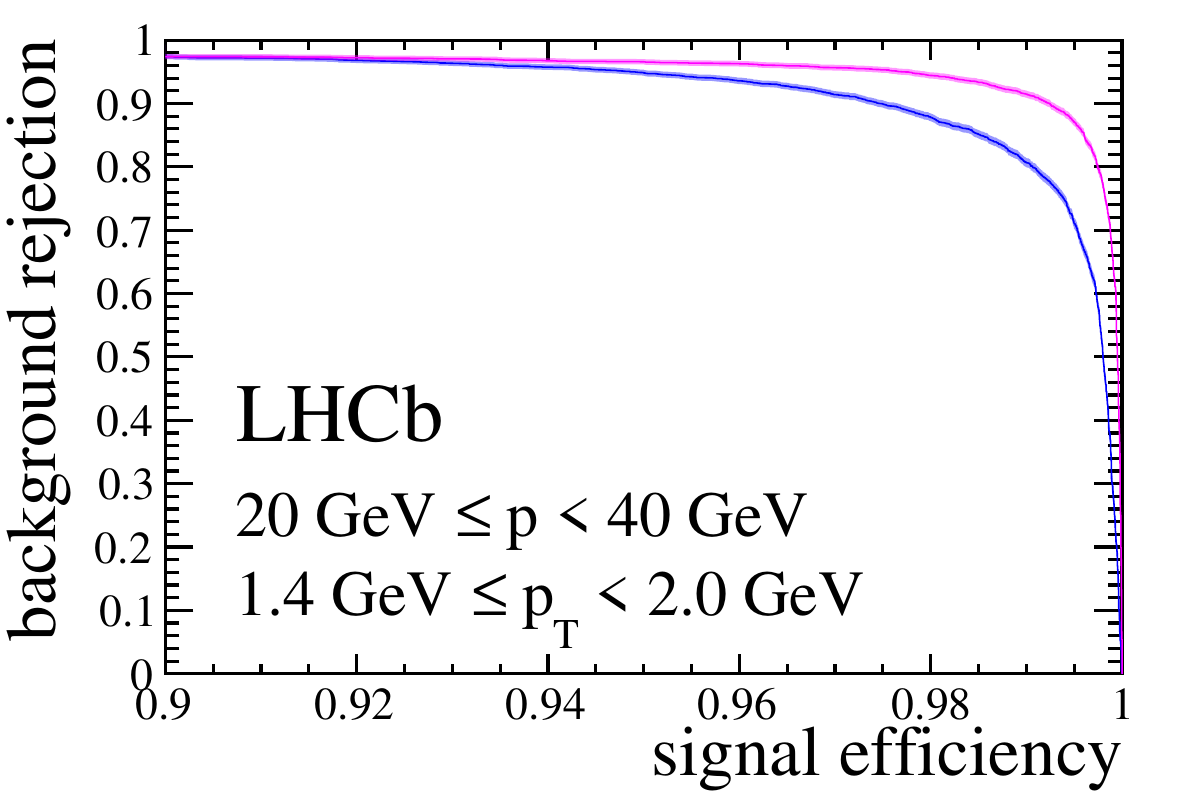}
    \includegraphics[width=0.32\textwidth]{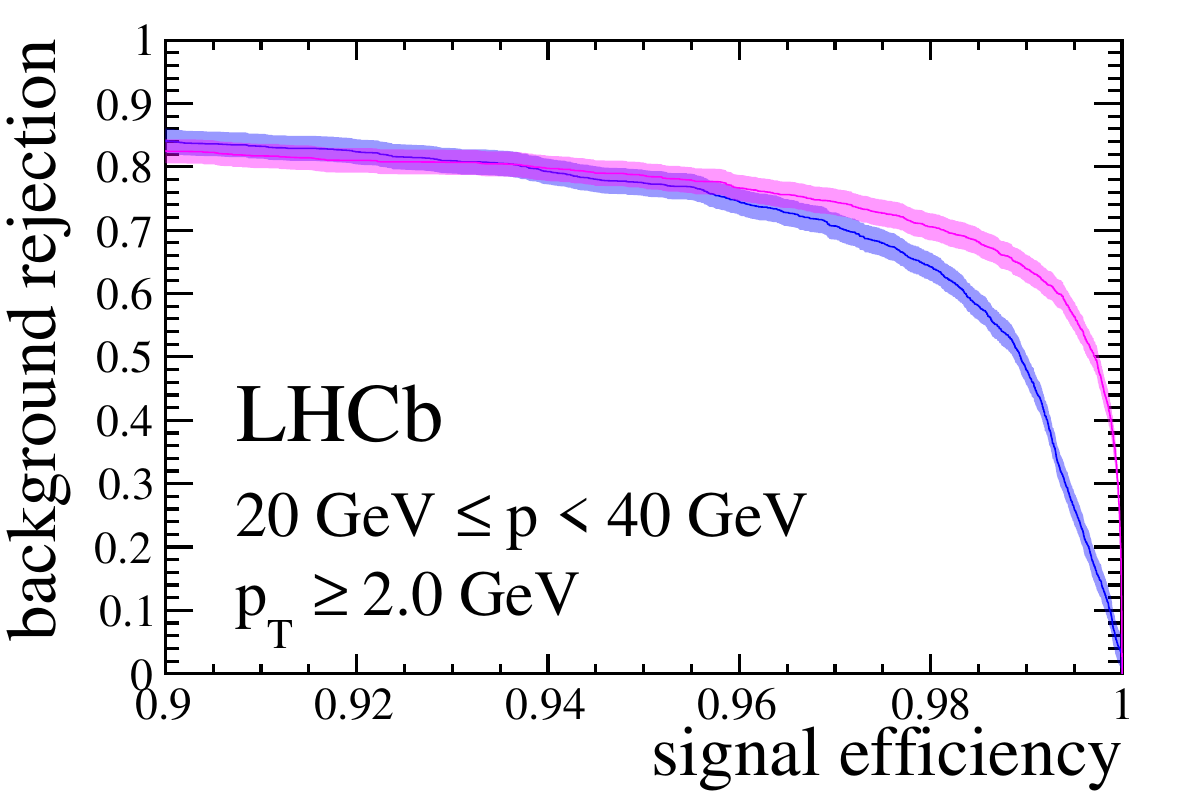}
    \includegraphics[width=0.32\textwidth]{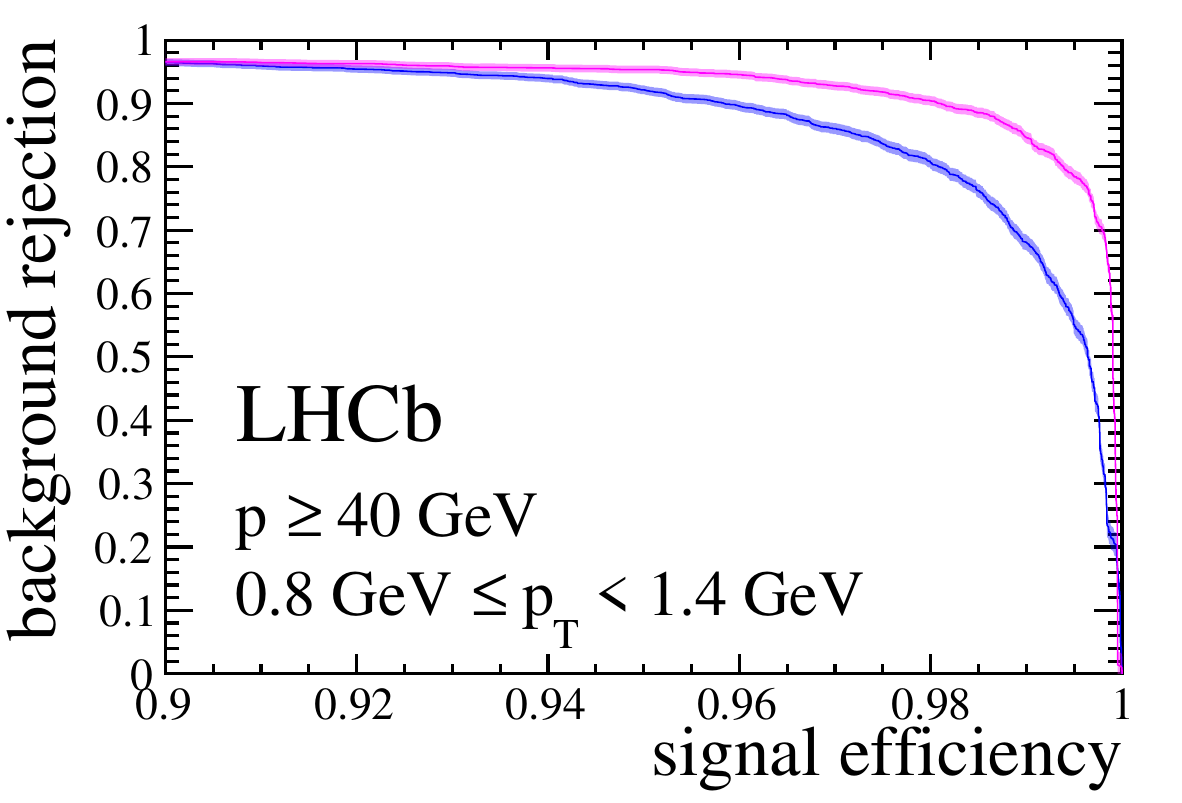}
    \includegraphics[width=0.32\textwidth]{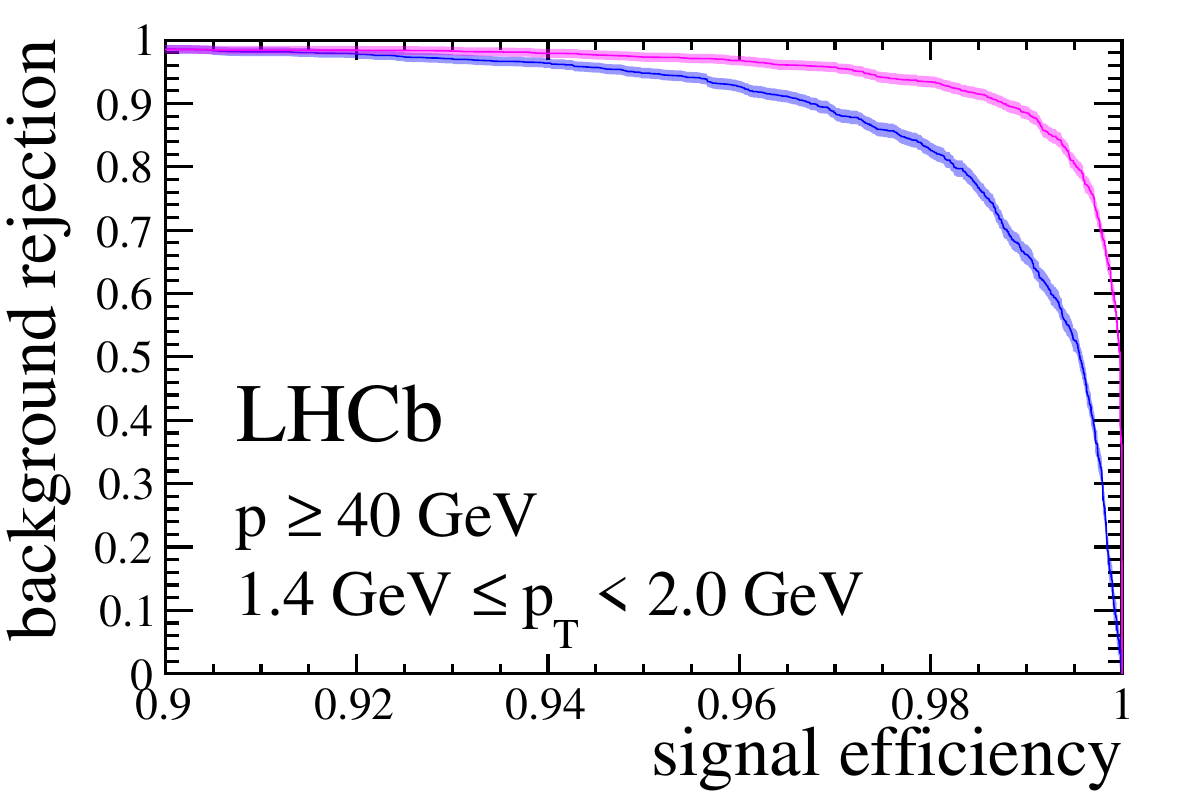}
    \includegraphics[width=0.32\textwidth]{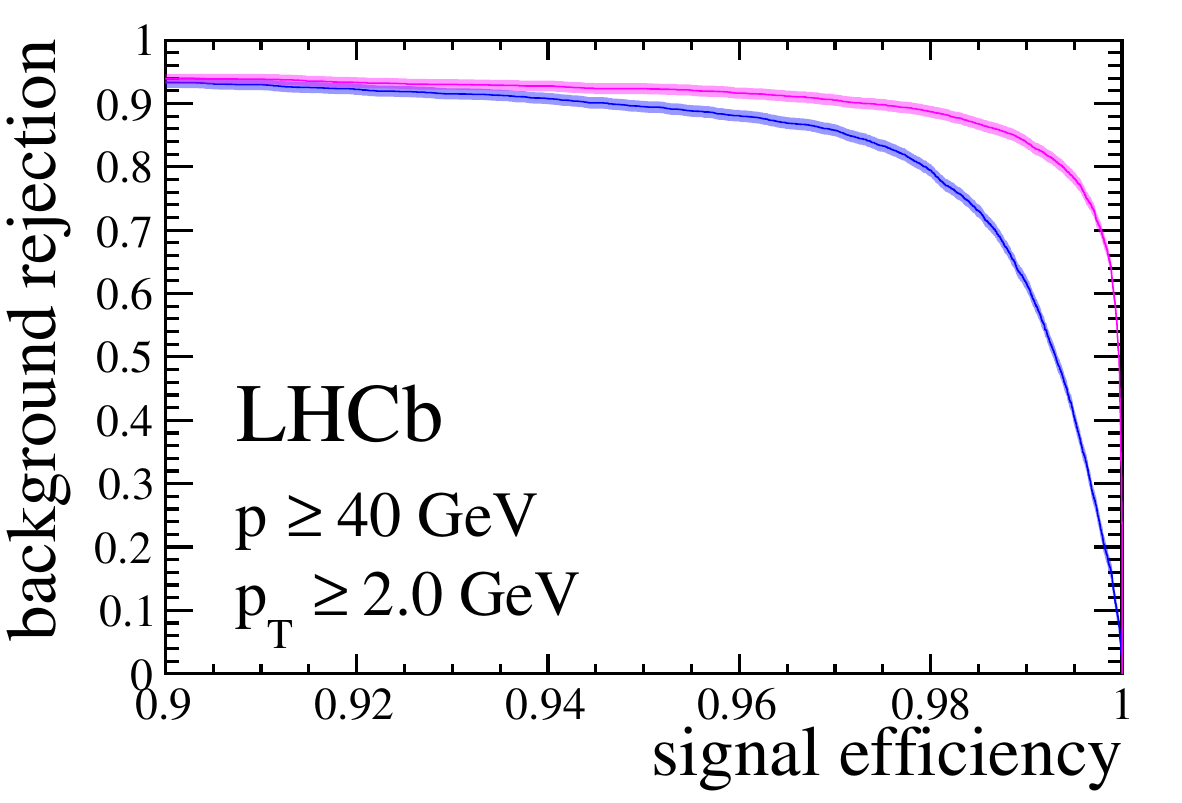}
    \caption{Proton rejection as a function of muon efficiency for tracks satisfying \texttt{IsMuon} obtained with the CatBoost algorithm (magenta) and \chicor (blue) on 2016
      calibration data. Low momentum bins, which are not covered by the calibration samples,
      are not shown. The bands represent statistical uncertainties.}
    \label{fig:runIIIp}
\end{figure}

Finally, since the classifier is trained on the same 2016 calibration data which are used to evaluate its performance, a cross-validation
method is used to obtain unbiased predictions. The dataset is split into $5$ subsets of
equal size, and the model is independently trained on all subsets but the $i$-th, for which predictions are made.
The ROC curves of the CatBoost algorithm are
shown in \Figref{fig:runIIIp} for muon efficiencies above $90\%$. For comparison, the ROC curves for the \chicor variable are superimposed.

As a result, for high muon efficiency the CatBoost algorithm has better discriminating power than \chicor in all the momentum bins. The greatest difference in background rejection is observed around 98\% muon efficiency, where for $p<10\,\gevc $ it lies in 20 -- 40\% range and 4 -- 10\% for $p\geq 10\,\gevc $.

With a similar setup as the one for \hltone (Sec.~\ref{ssec:perf}), throughput tests are performed on simulated Run 3 data and show a resource usage of about 0.4\% out of a total \hlttwo throughput rate of $\sim 129$ Hz. 
Therefore, the CatBoost operator is fast enough to be employed in the upgraded \hlttwo trigger of the experiment.\footnote{Being around 10 times slower than \chicor, the CatBoost implementation is not currently cost effective for \hltone.}

\section{Conclusions}
\label{sec:conclusions}

Two new muon identification algorithms have been developed in view of the \lhcb Run 3 upgrade. The first one, \chicor, expands on the muon likelihood variable developed in Run 1 by including the correlation among the muon hits. The second one features a multivariate algorithm based on the CatBoost
machine learning toolkit.
The performances of both algorithms in
terms of background rejection versus signal efficiency are characterised on 2016 proton calibration data, and in both cases  
are found to improve considerably those of the muon likelihood used during Run 1 and Run 2, with the CatBoost classifier offering a slightly better performance.
As far as the computational times are concerned, the \chicor has been proven to be fast enough to be included in the upgrade \hltone muon trigger lines.
The CatBoost algorithm, while improving the competing state-of-art gradient boosting libraries, can be computed
in the \hlttwo, where the time constraints are less stringent.

\section*{Acknowledgements}
%
% These Acknowledgements valid from 3-May-2019
%
\noindent 
We would like to thank the LHCb RTA team for supporting this publication, and in particular Vladimir Gligorov for the review and Manuel Schiller for the code optimisation.
We express our gratitude to our colleagues in the CERN
accelerator departments for the excellent performance of the LHC. We
thank the technical and administrative staff at the LHCb
institutes.
We acknowledge support from CERN and from the national agencies:
CAPES, CNPq, FAPERJ and FINEP (Brazil); 
MOST and NSFC (China); 
CNRS/IN2P3 (France); 
BMBF, DFG and MPG (Germany); 
INFN (Italy); 
NWO (Netherlands); 
MNiSW and NCN (Poland); 
MEN/IFA (Romania); 
MSHE (Russia); 
MinECo (Spain); 
SNSF and SER (Switzerland); 
NASU (Ukraine); 
STFC (United Kingdom); 
DOE NP and NSF (USA).
We acknowledge the computing resources that are provided by CERN, IN2P3
(France), KIT and DESY (Germany), INFN (Italy), SURF (Netherlands),
PIC (Spain), GridPP (United Kingdom), RRCKI, Yandex
LLC, HPC facilities at NRU HSE (Russia), CSCS (Switzerland), IFIN-HH (Romania), CBPF (Brazil),
PL-GRID (Poland) and OSC (USA).
We are indebted to the communities behind the multiple open-source
software packages on which we depend.
Individual groups or members have received support from
AvH Foundation (Germany);
EPLANET, Marie Sk\l{}odowska-Curie Actions and ERC (European Union);
ANR, Labex P2IO and OCEVU, and R\'{e}gion Auvergne-Rh\^{o}ne-Alpes (France);
Key Research Program of Frontier Sciences of CAS, CAS PIFI, and the Thousand Talents Program (China);
Basic Research Program of the NRU HSE and Yandex LLC (Russia);
GVA, XuntaGal and GENCAT (Spain);
the Royal Society
and the Leverhulme Trust (United Kingdom).

\newpage

\addcontentsline{toc}{section}{References}
\setboolean{inbibliography}{true}
\bibliographystyle{LHCb}
\bibliography{main,LHCb-PAPER,LHCb-CONF,LHCb-DP,LHCb-TDR}

\end{document}